\documentclass[sn-mathphys,Numbered]{sn-jnl}

\usepackage{graphicx}%
\usepackage{multirow}%
\usepackage{amsmath,amssymb,amsfonts}%
\usepackage{amsthm}%
\usepackage{mathrsfs}%
\usepackage[title]{appendix}%
\usepackage{xcolor}%
\usepackage{textcomp}%
\usepackage{manyfoot}%
\usepackage{booktabs}%
\usepackage{algorithm}%
\usepackage{algorithmicx}%
\usepackage{algpseudocode}%
\usepackage{listings}%
\usepackage{pgfplots}
\usepackage{pgf,tikz}
\usepackage{caption}
\usepackage{subcaption}
\usepackage{color}

\theoremstyle{thmstyleone}%
\theoremstyle{thmstyletwo}%
\newcommand{\Z}{\mathbb{Z}} 

\theoremstyle{thmstylethree}%

\pgfplotsset{compat=1.13}
\usepgfplotslibrary{fillbetween}
\usetikzlibrary{patterns}
\usetikzlibrary{arrows}

\begin{document}

\title[Lattice Fundamental Measure Theory beyond 0D Cavities: Dimers on Square Lattices]{Lattice Fundamental Measure Theory beyond 0D Cavities: Dimers on Square Lattices}


\author*[1]{\fnm{Michael} \sur{Zimmermann}}\email{mi.zimmermann@uni-tuebingen.de}
\equalcont{These authors contributed equally to this work.}

\author[1]{\fnm{Martin} \sur{Oettel}}\email{martin.oettel@uni-tuebingen.de}
\equalcont{These authors contributed equally to this work.}

\affil*[1]{\orgdiv{Institut für Angewandte Physik}, \orgname{Universität Tübingen}, \orgaddress{\street{Auf der Morgenstelle 10}, \city{Tübingen}, \postcode{72076}, \country{Germany}}}

\abstract{Using classical density functional theory, we study the behavior of dimers, i.e. hard rods of length $L=2$, on a two-dimensional cubic lattice. 
For deriving a free energy functional, we employ Levy’s prescription which is based on the minimization of a microscopic free energy with respect to the many-body probability under the constraint of a fixed density profile. Using that, we recover the functional originally found by Lafuente and Cuesta and derive an extension.
With this extension, the free energy functional is exact on cavities that can hold at most two particles simultaneously. 
The new functional improves the prediction of the free energy in bulk as well as in highly confined systems, especially for high packing fractions, in comparison to that of Lafuente and Cuesta.}




\maketitle

\section{Introduction}\label{sec1}
\newcommand{\allgemein}{%
\begin{tikzpicture}[line cap=round,line join=round,>=triangle 45,x=1cm,y=1cm]
\begin{scriptsize}
\draw [fill=blue] (0.5,1.55) circle (4pt);
\draw [fill=blue] (1.2,1.55) circle (4pt);
\draw [fill=blue] (1.55,1.55) circle (4pt);
\draw [color=red,line width=1pt] (1.55,0.5)-- ++(-4pt,-4pt) -- ++(8pt,8pt) ++(-8pt,0) -- ++(8pt,-8pt);
\draw [color=red,line width=1pt] (1.55,1.2)-- ++(-4pt,-4pt) -- ++(8pt,8pt) ++(-8pt,0) -- ++(8pt,-8pt);
\draw [color=red,line width=1pt] (1.55,1.55)-- ++(-4pt,-4pt) -- ++(8pt,8pt) ++(-8pt,0) -- ++(8pt,-8pt);
\draw [fill=black] (0.75,1.55) circle (0.5pt);
\draw [fill=black] (0.85,1.55) circle (0.5pt);
\draw [fill=black] (0.95,1.55) circle (0.5pt);
\draw [fill=black] (1.55,0.75) circle (0.5pt);
\draw [fill=black] (1.55,0.85) circle (0.5pt);
\draw [fill=black] (1.55,0.95) circle (0.5pt);
\end{scriptsize}
\end{tikzpicture}%

} 

\newcommand{\kreuz}{
\begin{tikzpicture}[line cap=round,line join=round,>=triangle 45,x=1cm,y=1cm,baseline=0.745ex]
\draw [line width=1pt] (-0.1,0.3)-- (0.1,0.3);
\draw [line width=1pt] (0.1,0.3)-- (0.1,0.5);
\draw [line width=1pt] (0.1,0.5)-- (0.3,0.5);
\draw [line width=1pt] (0.3,0.5)-- (0.3,0.3);
\draw [line width=1pt] (0.3,0.3)-- (0.5,0.3);
\draw [line width=1pt] (0.5,0.3)-- (0.5,0.1);
\draw [line width=1pt] (0.5,0.1)-- (0.3,0.1);
\draw [line width=1pt] (0.3,0.1)-- (0.3,-0.1);
\draw [line width=1pt] (0.3,-0.1)-- (0.1,-0.1);
\draw [line width=1pt] (0.1,-0.1)-- (0.1,0.1);
\draw [line width=1pt] (0.1,0.1)-- (-0.1,0.1);
\draw [line width=1pt] (-0.1,0.1)-- (-0.1,0.3);
\begin{scriptsize}
\draw [fill=blue] (0,0.2) circle (2pt);
\draw [fill=blue] (0.2,0.2) circle (2pt);
\draw [color=red,line width=1pt] (0.2,0) ++(-2pt,2pt) -- ++(4pt,-4pt) ++(0pt,4pt) -- ++(-4pt,-4pt) ;
\draw [color=red,line width=1pt]  (0.2,0.2) ++(-2pt,2pt) -- ++(4pt,-4pt) ++(0pt,4pt) -- ++(-4pt,-4pt);
\end{scriptsize}
\end{tikzpicture}
}

\newcommand{\kreuzeins}{
\begin{tikzpicture}[line cap=round,line join=round,>=triangle 45,x=1cm,y=1cm,baseline=0.745ex]
\draw [line width=1pt] (-0.1,0.3)-- (0.1,0.3);
\draw [line width=1pt] (0.1,0.3)-- (0.1,0.5);
\draw [line width=1pt] (0.1,0.5)-- (0.3,0.5);
\draw [line width=1pt] (0.3,0.5)-- (0.3,0.3);
\draw [line width=1pt] (0.3,0.3)-- (0.5,0.3);
\draw [line width=1pt] (0.5,0.3)-- (0.5,0.1);
\draw [line width=1pt] (0.5,0.1)-- (0.3,0.1);
\draw [line width=1pt] (0.3,0.1)-- (0.3,-0.1);
\draw [line width=1pt] (0.3,-0.1)-- (0.1,-0.1);
\draw [line width=1pt] (0.1,-0.1)-- (0.1,0.1);
\draw [line width=1pt] (0.1,0.1)-- (-0.1,0.1);
\draw [line width=1pt] (-0.1,0.1)-- (-0.1,0.3);
\begin{scriptsize}
\draw [fill=blue] (0,0.2) circle (2pt);
\end{scriptsize}
\end{tikzpicture}
}

\newcommand{\kreuzzwei}{
\begin{tikzpicture}[line cap=round,line join=round,>=triangle 45,x=1cm,y=1cm,baseline=0.745ex]
\draw [line width=1pt] (-0.1,0.3)-- (0.1,0.3);
\draw [line width=1pt] (0.1,0.3)-- (0.1,0.5);
\draw [line width=1pt] (0.1,0.5)-- (0.3,0.5);
\draw [line width=1pt] (0.3,0.5)-- (0.3,0.3);
\draw [line width=1pt] (0.3,0.3)-- (0.5,0.3);
\draw [line width=1pt] (0.5,0.3)-- (0.5,0.1);
\draw [line width=1pt] (0.5,0.1)-- (0.3,0.1);
\draw [line width=1pt] (0.3,0.1)-- (0.3,-0.1);
\draw [line width=1pt] (0.3,-0.1)-- (0.1,-0.1);
\draw [line width=1pt] (0.1,-0.1)-- (0.1,0.1);
\draw [line width=1pt] (0.1,0.1)-- (-0.1,0.1);
\draw [line width=1pt] (-0.1,0.1)-- (-0.1,0.3);
\begin{scriptsize}
\draw [fill=blue] (0.2,0.2) circle (2pt);
\end{scriptsize}
\end{tikzpicture}
}

\newcommand{\kreuzdrei}{
\begin{tikzpicture}[line cap=round,line join=round,>=triangle 45,x=1cm,y=1cm,baseline=0.745ex]
\draw [line width=1pt] (-0.1,0.3)-- (0.1,0.3);
\draw [line width=1pt] (0.1,0.3)-- (0.1,0.5);
\draw [line width=1pt] (0.1,0.5)-- (0.3,0.5);
\draw [line width=1pt] (0.3,0.5)-- (0.3,0.3);
\draw [line width=1pt] (0.3,0.3)-- (0.5,0.3);
\draw [line width=1pt] (0.5,0.3)-- (0.5,0.1);
\draw [line width=1pt] (0.5,0.1)-- (0.3,0.1);
\draw [line width=1pt] (0.3,0.1)-- (0.3,-0.1);
\draw [line width=1pt] (0.3,-0.1)-- (0.1,-0.1);
\draw [line width=1pt] (0.1,-0.1)-- (0.1,0.1);
\draw [line width=1pt] (0.1,0.1)-- (-0.1,0.1);
\draw [line width=1pt] (-0.1,0.1)-- (-0.1,0.3);
\begin{scriptsize}
\draw [color=red,line width=1pt] (0.2,0) ++(-2pt,2pt) -- ++(4pt,-4pt) ++(0pt,4pt) -- ++(-4pt,-4pt) ;
\end{scriptsize}
\end{tikzpicture}
}

\newcommand{\kreuzvier}{
\begin{tikzpicture}[line cap=round,line join=round,>=triangle 45,x=1cm,y=1cm,baseline=0.745ex]
\draw [line width=1pt] (-0.1,0.3)-- (0.1,0.3);
\draw [line width=1pt] (0.1,0.3)-- (0.1,0.5);
\draw [line width=1pt] (0.1,0.5)-- (0.3,0.5);
\draw [line width=1pt] (0.3,0.5)-- (0.3,0.3);
\draw [line width=1pt] (0.3,0.3)-- (0.5,0.3);
\draw [line width=1pt] (0.5,0.3)-- (0.5,0.1);
\draw [line width=1pt] (0.5,0.1)-- (0.3,0.1);
\draw [line width=1pt] (0.3,0.1)-- (0.3,-0.1);
\draw [line width=1pt] (0.3,-0.1)-- (0.1,-0.1);
\draw [line width=1pt] (0.1,-0.1)-- (0.1,0.1);
\draw [line width=1pt] (0.1,0.1)-- (-0.1,0.1);
\draw [line width=1pt] (-0.1,0.1)-- (-0.1,0.3);
\begin{scriptsize}
\draw [color=red,line width=1pt]  (0.2,0.2) ++(-2pt,2pt) -- ++(4pt,-4pt) ++(0pt,4pt) -- ++(-4pt,-4pt);
\end{scriptsize}
\end{tikzpicture}
}

\newcommand{\nkreuz}{%
\begin{tikzpicture}[line cap=round,line join=round,>=triangle 45,x=1cm,y=1cm, baseline=0.0ex]
\begin{scriptsize}
\draw [fill=blue] (0,0.2) circle (2pt);
\draw [fill=blue] (0.2,0.2) circle (2pt);
\draw [color=red,line width=1pt] (0.2,0) ++(-2pt,2pt) -- ++(4pt,-4pt) ++(0pt,4pt) -- ++(-4pt,-4pt) ;
\draw [color=red,line width=1pt]  (0.2,0.2) ++(-2pt,2pt) -- ++(4pt,-4pt) ++(0pt,4pt) -- ++(-4pt,-4pt);
\end{scriptsize}
\end{tikzpicture}%
} 

\newcommand{\nwaagrecht}{%
\begin{tikzpicture}[line cap=round,line join=round,>=triangle 45,x=1cm,y=1cm,baseline=1.74ex]
\begin{scriptsize}
\draw [fill=blue] (0.35,0.35) circle (2pt);
\end{scriptsize}
\end{tikzpicture}%
} 

\newcommand{\nsenkrecht}{%
\begin{tikzpicture}[line cap=round,line join=round,>=triangle 45,x=1cm,y=1cm,baseline=0.74675ex]
\begin{scriptsize}
\draw [color=red,line width=1pt]  (0.2,0.2) ++(-2pt,2pt) -- ++(4pt,-4pt) ++(0pt,4pt) -- ++(-4pt,-4pt);
\end{scriptsize}
\end{tikzpicture}%
} 

\newcommand{\quadrat}{
\begin{tikzpicture}[line cap=round,line join=round,>=triangle 45,x=1cm,y=1cm,baseline=0.0ex]
\draw [line width=1pt] (-0.15,-0.15)-- (-0.15,0.35);
\draw [line width=1pt] (-0.15,0.35)-- (0.35,0.35);
\draw [line width=1pt] (0.35,0.35)-- (0.35,-0.15);
\draw [line width=1pt] (0.35,-0.15)-- (-0.15,-0.15);
\begin{scriptsize}
\draw [fill=blue] (0,0) circle (2pt);
\draw [fill=blue] (0,0.2) circle (2pt);
\draw [color=red,line width=1pt] (0,0) ++(-2pt,2pt) -- ++(4pt,-4pt) ++(0pt,4pt) -- ++(-4pt,-4pt) ;
\draw [color=red,line width=1pt]  (0.2,0) ++(-2pt,2pt) -- ++(4pt,-4pt) ++(0pt,4pt) -- ++(-4pt,-4pt);
\end{scriptsize}
\end{tikzpicture}%
} 

\newcommand{\quadrateins}{
\begin{tikzpicture}[line cap=round,line join=round,>=triangle 45,x=1cm,y=1cm,baseline=0.0ex]
\draw [line width=1pt] (-0.15,-0.15)-- (-0.15,0.35);
\draw [line width=1pt] (-0.15,0.35)-- (0.35,0.35);
\draw [line width=1pt] (0.35,0.35)-- (0.35,-0.15);
\draw [line width=1pt] (0.35,-0.15)-- (-0.15,-0.15);
\begin{scriptsize}
\draw [fill=blue] (0,0) circle (2pt);
\draw [color=red,line width=1pt] (0,0) ++(-2pt,2pt) -- ++(4pt,-4pt) ++(0pt,4pt) -- ++(-4pt,-4pt) ;
\end{scriptsize}
\end{tikzpicture}%
} 

\newcommand{\quadratzwei}{
\begin{tikzpicture}[line cap=round,line join=round,>=triangle 45,x=1cm,y=1cm,baseline=0.0ex]
\draw [line width=1pt] (-0.15,-0.15)-- (-0.15,0.35);
\draw [line width=1pt] (-0.15,0.35)-- (0.35,0.35);
\draw [line width=1pt] (0.35,0.35)-- (0.35,-0.15);
\draw [line width=1pt] (0.35,-0.15)-- (-0.15,-0.15);
\begin{scriptsize}
\draw [fill=blue] (0,0) circle (2pt);
\draw [color=red,line width=1pt]  (0.2,0) ++(-2pt,2pt) -- ++(4pt,-4pt) ++(0pt,4pt) -- ++(-4pt,-4pt);
\end{scriptsize}
\end{tikzpicture}%
} 

\newcommand{\quadratdrei}{
\begin{tikzpicture}[line cap=round,line join=round,>=triangle 45,x=1cm,y=1cm,baseline=0.0ex]
\draw [line width=1pt] (-0.15,-0.15)-- (-0.15,0.35);
\draw [line width=1pt] (-0.15,0.35)-- (0.35,0.35);
\draw [line width=1pt] (0.35,0.35)-- (0.35,-0.15);
\draw [line width=1pt] (0.35,-0.15)-- (-0.15,-0.15);
\begin{scriptsize}
\draw [fill=blue] (0,0.2) circle (2pt);
\draw [color=red,line width=1pt] (0,0) ++(-2pt,2pt) -- ++(4pt,-4pt) ++(0pt,4pt) -- ++(-4pt,-4pt) ;
\end{scriptsize}
\end{tikzpicture}%
} 

\newcommand{\quadratvier}{
\begin{tikzpicture}[line cap=round,line join=round,>=triangle 45,x=1cm,y=1cm,baseline=0.0ex]
\draw [line width=1pt] (-0.15,-0.15)-- (-0.15,0.35);
\draw [line width=1pt] (-0.15,0.35)-- (0.35,0.35);
\draw [line width=1pt] (0.35,0.35)-- (0.35,-0.15);
\draw [line width=1pt] (0.35,-0.15)-- (-0.15,-0.15);
\begin{scriptsize}
\draw [fill=blue] (0,0.2) circle (2pt);
\draw [color=red,line width=1pt]  (0.2,0) ++(-2pt,2pt) -- ++(4pt,-4pt) ++(0pt,4pt) -- ++(-4pt,-4pt);
\end{scriptsize}
\end{tikzpicture}%
} 

\newcommand{\quadratfuenf}{
\begin{tikzpicture}[line cap=round,line join=round,>=triangle 45,x=1cm,y=1cm,baseline=0.0ex]
\draw [line width=1pt] (-0.15,-0.15)-- (-0.15,0.35);
\draw [line width=1pt] (-0.15,0.35)-- (0.35,0.35);
\draw [line width=1pt] (0.35,0.35)-- (0.35,-0.15);
\draw [line width=1pt] (0.35,-0.15)-- (-0.15,-0.15);
\begin{scriptsize}
\draw [fill=blue] (0,0) circle (2pt);
\end{scriptsize}
\end{tikzpicture}%
} 

\newcommand{\quadratsechs}{
\begin{tikzpicture}[line cap=round,line join=round,>=triangle 45,x=1cm,y=1cm,baseline=0.0ex]
\draw [line width=1pt] (-0.15,-0.15)-- (-0.15,0.35);
\draw [line width=1pt] (-0.15,0.35)-- (0.35,0.35);
\draw [line width=1pt] (0.35,0.35)-- (0.35,-0.15);
\draw [line width=1pt] (0.35,-0.15)-- (-0.15,-0.15);
\begin{scriptsize}
\draw [fill=blue] (0,0.2) circle (2pt);
\end{scriptsize}
\end{tikzpicture}%
} 

\newcommand{\quadratsieben}{
\begin{tikzpicture}[line cap=round,line join=round,>=triangle 45,x=1cm,y=1cm,baseline=0.0ex]
\draw [line width=1pt] (-0.15,-0.15)-- (-0.15,0.35);
\draw [line width=1pt] (-0.15,0.35)-- (0.35,0.35);
\draw [line width=1pt] (0.35,0.35)-- (0.35,-0.15);
\draw [line width=1pt] (0.35,-0.15)-- (-0.15,-0.15);
\begin{scriptsize}
\draw [color=red,line width=1pt]  (0,0) ++(-2pt,2pt) -- ++(4pt,-4pt) ++(0pt,4pt) -- ++(-4pt,-4pt);
\end{scriptsize}
\end{tikzpicture}%
} 

\newcommand{\quadratacht}{
\begin{tikzpicture}[line cap=round,line join=round,>=triangle 45,x=1cm,y=1cm,baseline=0.0ex]
\draw [line width=1pt] (-0.15,-0.15)-- (-0.15,0.35);
\draw [line width=1pt] (-0.15,0.35)-- (0.35,0.35);
\draw [line width=1pt] (0.35,0.35)-- (0.35,-0.15);
\draw [line width=1pt] (0.35,-0.15)-- (-0.15,-0.15);
\begin{scriptsize}
\draw [color=red,line width=1pt]  (0.2,0) ++(-2pt,2pt) -- ++(4pt,-4pt) ++(0pt,4pt) -- ++(-4pt,-4pt);
\end{scriptsize}
\end{tikzpicture}%
} 

\newcommand{\nquadrat}{
\begin{tikzpicture}[line cap=round,line join=round,>=triangle 45,x=1cm,y=1cm,baseline=0.0ex]
\begin{scriptsize}
\draw [fill=blue] (0,0) circle (2pt);
\draw [fill=blue] (0,0.2) circle (2pt);
\draw [color=red,line width=1pt] (0,0) ++(-2pt,2pt) -- ++(4pt,-4pt) ++(0pt,4pt) -- ++(-4pt,-4pt) ;
\draw [color=red,line width=1pt]  (0.2,0) ++(-2pt,2pt) -- ++(4pt,-4pt) ++(0pt,4pt) -- ++(-4pt,-4pt);
\end{scriptsize}
\end{tikzpicture}%
} 

\newcommand{\nleins}{
\begin{tikzpicture}[line cap=round,line join=round,>=triangle 45,x=1cm,y=1cm,baseline=-0.575ex]
\begin{scriptsize}
\draw [fill=blue] (0,0) circle (2pt);
\draw [color=red,line width=1pt] (0,0) ++(-2pt,2pt) -- ++(4pt,-4pt) ++(0pt,4pt) -- ++(-4pt,-4pt) ;
\end{scriptsize}
\end{tikzpicture}%
} 

\newcommand{\nlzwei}{
\begin{tikzpicture}[line cap=round,line join=round,>=triangle 45,x=1cm,y=1cm,baseline=-0.575ex]
\begin{scriptsize}
\draw [fill=blue] (0,0) circle (2pt);
\draw [color=red,line width=1pt]  (0.2,0) ++(-2pt,2pt) -- ++(4pt,-4pt) ++(0pt,4pt) -- ++(-4pt,-4pt);
\end{scriptsize}
\end{tikzpicture}%
} 

\newcommand{\nldrei}{
\begin{tikzpicture}[line cap=round,line join=round,>=triangle 45,x=1cm,y=1cm,baseline=0.00ex]
\begin{scriptsize}
\draw [fill=blue] (0,0.2) circle (2pt);
\draw [color=red,line width=1pt] (0,0) ++(-2pt,2pt) -- ++(4pt,-4pt) ++(0pt,4pt) -- ++(-4pt,-4pt) ;
\end{scriptsize}
\end{tikzpicture}%
} 

\newcommand{\nlvier}{
\begin{tikzpicture}[line cap=round,line join=round,>=triangle 45,x=1cm,y=1cm,baseline=0.0ex]
\begin{scriptsize}
\draw [fill=blue] (0,0.2) circle (2pt);
\draw [color=red,line width=1pt]  (0.2,0) ++(-2pt,2pt) -- ++(4pt,-4pt) ++(0pt,4pt) -- ++(-4pt,-4pt);
\end{scriptsize}
\end{tikzpicture}%
} 

The interaction of $k$-mers as a model for polymers is a topic with a long historical background in statistical physics.
To simplify this model, $k$-mers were treated as hard rods and the continuous space was replaced by a lattice system.
For dimers, i.e. $k=2$, Fisher solved the so-called dimer problem in 1961 \cite{fisher61}. 
He found a method to quantify the possibilities of completely covering of a subvolume of $\Z^2$ by using only vertical and horizontal rods of length 2. 
This allowed him, for example, to find the free energy of this system for an equal number of horizontal and vertical dimers.
However, the determination of the free energy of arbitrary densities of dimers on a 2D lattice remained open.
In particular, for inhomogeneous, lattice point dependent densities, there was no good handle on the associated free energy functional
until Lafuente and Cuesta (abbreviated as LC) established a method based on the exact Fundamental Measure Theory for hard rods on a 1D system \cite{lafuente02,lafuente03}. 
Functionals in higher dimensions (termed LC functionals in this paper) could be approximated using the exact 1D result, similar to the dimensional crossover introduced by Tarazona and Rosenfeld for hard sphere systems in continuous space \cite{tarazona97,tarazona00}.
Their result allows to treat not only hard rods of one species, but also mixtures of them. 
To the best of our knowledge, there have been no further improvements to the LC functionals reported in the literature, and it is the aim of the paper
to construct an extension of the LC functional for the specific case of dimers on a 2D lattice system. Similar to LC, the starting point are free energies for extremely confined situations: cavities which can hold only a few particles. The results are generalized to any inhomogeneous situation using a subtraction method as introduced by LC. 
Whereas the LC functional gives the exact free energy for any cavity that holds one particle, the extended functional gives the exact free energy also
for cavities with two particles.

There are several ways to introduce density functional for classical systems. 
A standard reference is Evans \cite{evans78} which can be viewed as a translation of the quantum mechanical results by Mermin into classical statistical mechanics \cite{evans92,mermin65}.
Similarly, Levy's construction of quantum density functional theory \cite{levy79} was translated to the classical realm by Dwandaru and Schmidt  
\cite{schmidt11}, which is presented in the following for lattice system.
The system of interest consists of hard rods on a finite subset $V$ of the 2D lattice $\Z^2$. 
On this volume only horizontal and vertical rods are considered, labeled with $o$ and $x$ respectively.
In the following, elements of $\Z^2$ are either described by $\textbf{s}$ or by $(s_1,s_2)$.
The aim of density functional theory is to describe the free energy by a functional of the one particle density for the orientation $k\in\{o,x\}$ given by
\begin{equation}
\rho^{k}(\textbf{s})=\text{Tr}_{\text{cl}} f \hat{\rho}^{k,N_k}(\textbf{s}),\label{Eq_constraint}
\end{equation} 
where the classical trace is $\text{Tr}_{\text{cl}}=\sum_{N_o,N_x=0}^{\infty} \frac{1}{N_o!N_x!}\sum_{\textbf{s}_1,\dots,\textbf{s}_{N_o+N_x}\in V}$ and $f=f^{N_o,N_x }(\textbf{s}_1;\dots ;\textbf{s}_{N_o+N_x})$ is the probability of finding $N_o$ horizontal particles, where the first one is placed at $\textbf{s}_1$, the second one at $\textbf{s}_2$ and so on up to $\textbf{s}_{N_o}$ and the vertical particles are placed at $\textbf{s}_{N_o+1}$ to $\textbf{s}_{N_o+N_x}$. 
Furthermore, the density operator
\begin{equation}
\hat{\rho}^{k,N_k}(\textbf{s})=\sum_{i=1}^{N_k} \delta\left(\textbf{s}-\textbf{s}_{i+\delta_{k,x}N_o}\right)
\end{equation}
is introduced, where $\delta_{k,x}$ is the Kronecker delta. 
Levy's functional is defined as
\begin{equation}
\mathcal{F}_{\text{L}}\left[\rho^o,\rho^x\right]=\min_{f\to \rho^o,\rho^x}\left[\text{Tr}_{\text{cl}} f \left(U+\beta^{-1}\log\left(f\right)\right)\right]
\end{equation}
and is exact for all one particle densities representable by Eq.~\ref{Eq_constraint} \cite{levy79}.
Here, $\beta=(k_{B}T)^{-1}$ describes the inverse of the product of the Boltzmann factor $k_B$ and the temperature $T$. Without loss of generality we assume $\beta=1$. 
Furthermore, $U$ is the interaction potential of the hard rods and the minimization is restricted to the set $\{f \text{ is probability density}: \rho^k=\text{Tr}_{\text{cl}} f \hat{\rho}^{k,N_k},k\in\{o,x\}\}$. 
The great advantage of working with Levy's functional on finite lattices compared to working on a continuous space is that all possible particle configurations can be described and enumerated easily. 
Consequently, at least numerically, the free energy for any configuration can be calculated.
Nevertheless, the set of free parameters to be minimized in Levy's description, even for relatively small volumes, becomes numerous.
Therefore, an analytical approximation to the correct solution is sought.

The paper is organized as follows. In section \ref{sec2}, the exact free energy on $0$D cavities, i.e. cavities that can hold  at most one particle simultaneously, for dimers is rederived using a Levy functional argument. 
Based on this result, the method of Lafuente and Cuesta to derive a free energy density and consequently a free energy functional is introduced.
A new free energy density is derived in section \ref{sec3} by minimizing Levy's functional on a larger cavity and combining this result with Lafuente and Cuesta's method.
In section \ref{sec4}, the new free energy functional is compared with the LC functional for the cases of a bulk system and a highly confined system.
Finally, in section \ref{sec5}, we conclude by relating the newly found result to previous findings for the dimer systems on a 2D lattice.

\section{Derivation of the Lattice FMT Functional for Dimers using Levy's Functional\label{sec2}}
\begin{figure}
\centering
\includegraphics[width=\textwidth]{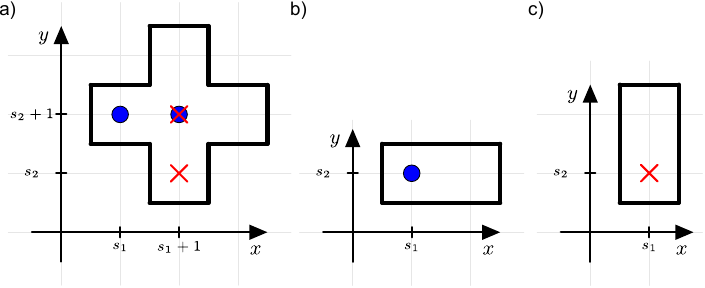}
\caption{Possible 0D cavities of a rod of length $L=2$ in two dimensions. The blue dots indicate possible locations for vertical rods and the red cross for horizontal rods. The cavities in (b) and (c) can be embedded in the maximal 0D cavity shown in (a). \label{LC_cavity}}
\end{figure}

First, by restricting the set of possible probability densities to those configurations where none of the particles overlap (i.e. $\text{Tr}_{\text{cl}} f U=0$), Levy's functional reduces to
\begin{equation}
\mathcal{F}_{\text{L}}\left[\rho^o,\rho^x\right]=\min_{f\to \rho}\left[\text{Tr}_{\text{cl}} f \beta^{-1}\log\left(f\right)\right].
\end{equation}
To further simplify the functional, we consider that the energy functional is invariant under permutation of the particles of the same species. The same must hold for the probability density $f_{\text{min}}$, which minimizes this functional. The definition
\begin{align}
&\psi^{(N_o,N_x)}\left(\textbf{s}_1;\dots; \textbf{s}_{N_o};\textbf{s}_{N_o+1};\dots; \textbf{s}_{N_o+N_x}\right)\nonumber \\
&:=\frac{1}{N_o!N_x!}\sum_{\sigma_o\in \Sigma_{N_o}}\sum_{\sigma_x\in \Sigma_{N_x}}f^{N_o,N_x}\left(\textbf{s}_{\sigma_o(1)};\dots;\textbf{s}_{\sigma_o(N_o)};\textbf{s}_{N_o+\sigma_x(1)};\dots; \textbf{s}_{N_o+\sigma_x(N_x)}\right),
\end{align}
where the sums run over all permutations of the length $N_o$ and $N_x$ respectively,
yields the final version of Levy's free energy functional for hard rods on lattices
\begin{equation}
\mathcal{F}_{\text{L}}[\rho^o,\rho^x]=\min_{\psi\to\rho}\sum_{\psi}\psi\log(\psi), \label{Eq_Levy}
\end{equation}
which will be used in the following.
For simplicity of the notation, we will drop the index ``min'', even if the probabilities are the one gained by minimizing Levy's functional.  
From here on, we restrict the theory to dimers, i.e. $L=2$, but most of the theory can be 
extended to larger rod sizes. The simplest connected volumes for which Levy's functional determines the free energy, are those that can at most hold one particle at a time. These volumes are called 0D cavities and for two dimensions some of them are shown in Fig.~\ref{LC_cavity} \cite{lafuente02}. They also can be defined as partial volumes of the maximal 0D cavity shown in Fig.~\ref{LC_cavity}.a.

By definition of the 0D cavity, all probabilities for $N_o+N_x>1$ vanish. 
The constraint in Levy's functional becomes trivial and yields for the probabilities of one particle in the volume of the maximal 0D cavity 
\begin{align}
\rho^{o}(s_1,s_2+1)&=\psi^{(1,0)}(s_1,s_2+1),\nonumber\\
\rho^{o}(s_1+1,s_2+1)&=\psi^{(1,0)}(s_1+1,s_2+1),\nonumber\\
\rho^{x}(s_1+1,s_2)&=\psi^{(0,1)}(s_1+1,s_2),\nonumber\\
\rho^{x}(s_1+1,s_2+1)&=\psi^{(0,1)}(s_1+1,s_2+1).
\end{align}
The normalization of $f$ also implies a normalization for $\psi$ (i.e. $\sum_\psi \psi=1$) and consequently the probability of the empty configuration is given by
\begin{align}
\psi^{(0,0)} &=1-\psi^{(1,0)}(s_1,s_2+1)-\psi^{(1,0)}(s_1+1,s_2+1)\nonumber\\
&-\psi^{(0,1)}(s_1+1,s_2)-\psi^{(0,1)}(s_1+1,s_2+1)\nonumber \\
&=1-\rho^{o}(s_1,s_2+1)-\rho^{o}(s_1,s_2+1)-\rho^{x}(s_1+1,s_2)-\rho^{x}(s_1,s_2+1). 
\end{align}
Therefore, with this constraint all parameters in Eq.~\ref{Eq_Levy} are defined and there are none left to be minimized. 
This results in a free energy
\begin{align}
\mathcal{F}_{\text{L}}^{0\text{D}}&=\sum_\psi \psi \log(\psi)\nonumber\\
&=\mathcal{F}_{\text{id}}^{0\text{D}}+\mathcal{F}_{\text{ex}}^{0\text{D}},\label{Eq_free_energy_levy}
\end{align}
where we split the free energy functional into an ideal gas part
\begin{align}
\mathcal{F}_{\text{id}}=\sum_{\textbf{s}} \sum_{k \in {o,x}}\rho^k\left(\textbf{s}\right)\log\left(\rho^k \left(\textbf{s}\right)\right)-\rho^k \left(\textbf{s}\right),\label{Eq_ideal_free_functional}
\end{align}
which is labelled by $0$D in Eq.~\ref{Eq_free_energy_levy} to indicate the evaluation within the maximal $0$D cavity and an excess part 
\begin{align*}
\mathcal{F}_{\text{ex}}^{0\text{D}}&=\psi^{(0,0)}\log\left(\psi^{(0,0)}\right)+\rho^{o}\left(s_1,s_2+1\right)+\rho^{o}\left(s_1+1,s_2+1\right)\\
&+\rho^{x}\left(s_1+1,s_2\right)+\rho^{x}\left(s_1+1,s_2+1\right)\\
&=:\Phi^{0\text{D}}\left(\rho^{o}\left(s_1,s_2+1\right),\rho^{o}\left(s_1+1,s_2+1\right),\rho^{x}\left(s_1+1,s_2\right),\rho^{x}\left(s_1+1,s_2+1\right)\right)
\end{align*}
as it is introduced from Lafuente and Cuesta \cite{lafuente02}.
Here, the function 
\begin{align}
&\Phi^{0\text{D}}\left(\rho^{o}_1,\dots,\rho^{o}_{N_o},\rho^{x}_1,\dots,\rho^{x}_{N_x}\right)\nonumber\\
&=\left(1-\sum_{k\in\{o,x\}}\sum_{j=1}^{n_k}\rho^{k}_j\right)\log\left(1-\sum_{k\in\{o,x\}}\sum_{j=1}^{n_k}\rho^{k}_j\right)+\sum_{k\in\{o,x\}}\sum_{j=1}^{n_k}\rho^{k}_j
\end{align}
 is used. 
For simplicity we introduce the notations 
\begin{align}
\nkreuz\left(s_1,s_2\right)&=\left(\rho^{o}\left(s_1,s_2+1\right),\rho^{o}\left(s_1+1,s_2+1\right),\rho^{x}\left(s_1+1,s_2\right),\rho^{x}\left(s_1+1,s_2+1\right)\right),\nonumber\\
\nwaagrecht\left(s_1,s_2\right)&=\left(\rho^{o}\left(s_1,s_2\right)\right),\nonumber\\
\nsenkrecht\left(s_1,s_2\right)&=\left(\rho^{x}\left(s_1,s_2\right)\right)\label{Eq_cavities_I}
\end{align}
similar to the notation of Lafuente and Cuesta in \cite{lafuente03}. In the following, we will drop the explicit notation of the evaluation point $\textbf{s}=(s_1,s_2)$, nevertheless, $\nkreuz$, $\nwaagrecht$ and $\nsenkrecht$ should be understood as functions depending on $\textbf{s}$.
Usually, in the classical fundamental measure density theory the sums over the densities in $\nkreuz$, $\nwaagrecht$ and $\nsenkrecht$ are interpreted as weighted densities, as in LC's work \cite{lafuente03} and in Rosenfeld's original work on fundamental measure theory in a continuous system \cite{rosenfeld89},
and in the evaluation of $\Phi^{0\text{D}}$ these weighted densities appear naturally.  
Nevertheless, we stick to our notation, because it can be generalized more easily in the following section, where it is found that the improved free energy density is not a function of simple weighted densities anymore. 

As a next step, we follow Lafuente and Cuesta's iterative scheme for finding an approximation for the free energy functional for an arbitrary subset of $\Z^2$ \cite{lafuente03}. 
The excess free energy is written as a sum over lattice points of an excess free energy density,
\begin{align}
 \mathcal{F}_{\text{ex}} = \sum_{(s_1,s_2)\in \Z^2} f^{\text{ex}}(\textbf{s})\;, 
\end{align}
and the first ansatz chosen by Lafuente and Cuesta is $f^{\text{ex}}(\textbf{s})=\Phi^{0\text{D}}\left(\nkreuz\right)$.
To be acceptable, the evaluation for arbitrary $0$D cavities should be exact. It suffices to consider the maximal $0$D cavity,
and the free energy for the maximal cavity evaluated with this first ansatz for the free energy density gives 
\begin{align*}
\left.\sum_{(s_1,s_2)\in \Z^2} f^{\text{ex}} \right|_{\kreuz}&=\Phi^{0\text{D}}\left(\kreuz\right)+\Phi^{0\text{D}}\left(\kreuzeins\right)+\Phi^{0\text{D}}\left(\kreuzzwei\right)\\
&+\Phi^{0\text{D}}\left(\kreuzdrei\right)+\Phi^{0\text{D}}\left(\kreuzvier\right),
\end{align*}
where the cross-like frame signifies an evaluation for a 0D cavity fixed at some point in $\Z^2$.
The unwanted additional terms beyond the first one on the rhs can be eliminated with a second ansatz for the free energy denisty, obtained by subtracting $\Phi^{0\text{D}}\left(\nwaagrecht\right)+\Phi^{0\text{D}}\left(\nsenkrecht\right)$ from the first ansatz. Therefore,
\begin{align}
f^{\text{ex}}=\Phi^{0\text{D}}\left(\nkreuz\right)-\Phi^{0\text{D}}\left(\nwaagrecht\right)-\Phi^{0\text{D}}\left(\nsenkrecht\right)\label{Eq_density_LC}
\end{align}
is an excess free energy density, which determines the free energy of an arbitrary 0D cavity correctly. 
This result was already found by Lafuente and Cuesta in \cite{lafuente02}, but Levy's formula provides an alternative elegant argument why the free energy of the 0D cavity 
has the form described above. 
Beyond this specific example, Lafuente and Cuesta proved that this subtraction method to obtain a free energy functional exact for 0D cavities works for arbitrary hard particles on arbitrary lattices \cite{lafuente05, lafuente04} and, as examples, they derived explicit functionals for hard (hyper-)cubes on 
square, triangular, simple cubic, face-centered cubic and body-centered lattices \cite{lafuente03_2}.

\section{An extension to lattice FMT for dimers}\label{sec3}
\subsection{The square cavity}
In contrast to Lafuente and Cuesta's method of finding their solution for the 0D cavity which is derived from the one dimensional exact solution for the FMT, Levy's formula is not restricted to 0D cavities.

\begin{figure}
\centering
\includegraphics{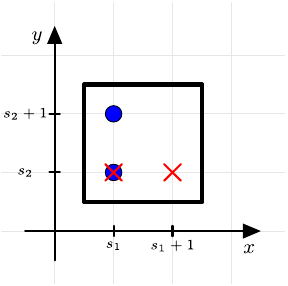}
\caption{A cavity, which can at most hold two vertical rods or two horizontal particles simultaneously.  The blue dots indicate possible locations for vertical rods and the red cross for horizontal rods. \label{squarecavity}}
\end{figure}

In Eq.~\ref{Eq_Levy}, the free energy depends on the specific probabilities $\psi^{(N_o,N_x)}$ for finding exactly $N_o$ horizontal and $N_x$ vertical rods.
These are linked by the constraints of having certain fixed densities $\rho^{o}(\textbf{s})$ and $\rho^{x}(\textbf{s})$. 
The one particle density of a horizontal rod placed at the location $\textbf{s}$ can be written as 
\begin{align}
\rho^{o}(\textbf{s})=\sum_{N_o\geq 1}\sum_{N_ x\geq 0}\sum_{(\textbf{s}_1;\dots;\textbf{s}_m)\in C^{N_o,N_x}_{o,\textbf{s}}}\psi^{(N_o,N_x)}\left(\textbf{s}_1;\dots;\textbf{s}_m\right);\label{Eq_constraint_I}
\end{align}
where $C_{o,\textbf{s}}^{N_o,N_x}$ is set of all allowed configurations (i.e. no overlapping of the rods) of $N_o$ horizontal and $N_x$ vertical rods, in which one horizontal rod is located at $\textbf{s}$. 
A similar constraint holds for the one particle density of vertical rods, namely
\begin{align}
\rho^{x}(\textbf{s})=\sum_{N_o\geq 0}\sum_{N_ x\geq 1}\sum_{(\textbf{s}_1;\dots;\textbf{s}_m)\in C_{x,\textbf{s}}^{N_o,N_x}}\psi^{(N_o,N_x)}(\textbf{s}_1;\dots;\textbf{s}_{m}).\label{Eq_constraint_II}
\end{align}
For the case of a square cavity shown in Fig.~\ref{squarecavity} (holding up to two particles),  these constraints  are no longer trivial as in the case of a 0D cavity, but yield:
\begin{align}
\rho^{o}\left(s_1,s_2\right)&=\psi^{(1,0)}\left(s_1,s_2\right)+\psi^{(2,0)}\left(s_1,s_2;s_1,s_2+1\right), \nonumber \\ 
\rho^{o}\left(s_1,s_2+1\right)&=\psi^{(1,0)}\left(s_1,s_2+1\right)+\psi^{(2,0)}\left(s_1,s_2;s_1,s_2+1\right), \nonumber\\
\rho^{x}\left(s_1,s_2\right)&=\psi^{(0,1)}\left(s_1,s_2\right)+\psi^{(0,2)}\left(s_1,s_2;s_1+1,s_2\right), \nonumber\\
\rho^{x}\left(s_1+1,s_2\right)&=\psi^{(0,1)}\left(s_1+1,s_2\right)+\psi^{(0,2)}\left(s_1,s_2;s_1+1,s_2\right). \label{Eq_constraint_square}
\end{align}
The interpretation of these equations is straightforward.
E.g., the one particle density for a horizontal hard rod at the location $\left(s_1,s_2\right)$, i.e. $\rho^{o}\left(s_1,s_2\right)$, is the sum over the probabilities 
of configurations with only one horizontal rod in the square cavity at $\left(s_1,s_2\right)$ and with two horizontal particles simultaneously are located at $(s_1,s_2)$ and at $(s_1,s_2+1)$.
For a simpler notation, we shorten $\psi^{(2,0)}(s_1,s_2;s_1,s_2+1)$ to $\psi^{(2,0)}$ and similarly $\psi^{(0,2)}(s_1,s_2;s_1+1,s_2)$ to $\psi^{(0,2)}$ and define the negative entropy as $S_\text{n}(\psi)=\psi\log(\psi)$.
The requirement that the sum of the probabilities of all possible configurations is one implies that 
\begin{align*}
\psi^{(0,0)}&=1-\sum_{\psi\neq \psi^{(0,0)}}\psi\\
&=1-\rho^{o}(s_1,s_2)-\rho^{o}(s_1,s_2+1)-\rho^{x}(s_1,s_2)-\rho^{x}(s_1+1,s_2)+\psi^{(2,0)}+\psi^{(0,2)}
\end{align*}
Therefore, after eliminating all $\psi^{(N_o,N_x)}$ with $N_o+N_x=1$ using Eqs. \ref{Eq_constraint_square},  Levy's formula can be expressed as follows
\begin{align}
&\mathcal{F}^{2\times2}_{\text{L}}=\min_{\psi\to\rho}\sum_{\psi}S_{\text{n}}(\psi)=\min_{\psi^{(2,0)},\psi^{(0,2)}} S_\text{n}\left(\psi^{(0,0)}\right)\nonumber\\
&+S_\text{n}\left(\rho^{o}(s_1,s_2)-\psi^{(2,0)}\right)+S_\text{n}\left(\rho^{o}(s_1,s_2+1)-\psi^{(2,0)}\right)+S_\text{n}\left(\rho^{x}(s_1,s_2)-\psi^{(0,2)}\right)\nonumber\\
&+S_\text{n}\left(\rho^{x}(s_1,s_2+1)-\psi^{(0,2)}\right)+S_\text{n}\left(\psi^{(2,0)}\right)+S_\text{n}\left(\psi^{(0,2)}\right). \label{Eq_Levy_square}
\end{align}
Hence, the square cavity contains the two free parameters  $\psi^{(2,0)}$ and $\psi^{(0,2)}$ for the minimization. 
Minimizing with respect to those yields the equations
\begin{align*}
\psi^{(2,0)}\psi^{(0,0)}=\left(\rho^{o}(s_1,s_2)-\psi^{(2,0)}\right)\left(\rho^{o}(s_1,s_2+1)-\psi^{(2,0)}\right),\\
\psi^{(0,2)}\psi^{(0,0)}=\left(\rho^{x}(s_1,s_2)-\psi^{(0,2)}\right)\left(\rho^{x}(s_1+1,s_2)-\psi^{(0,2)}\right).
\end{align*}

This result can be generalized for arbitrary cavities.
The minimization of Levy's formula in Eq.~\ref{Eq_Levy} defines the dependency between the free parameters, i.e. the many particle probabilities $\psi^{(N_o,N_x)}$ with $N_o+N_x\geq 2$, and the probabilities for configuration with only one particle. In the general case, the connection is described by
\begin{align}
&\psi^{(N_o,N_x)}(\textbf{s}_1;\dots;\textbf{s}_{N_o};\textbf{s}_{N_o+1};\dots;\textbf{s}_{N_o+N_x})\left(\psi^{(0,0)}\right)^{(N_o+N_x-1)}\nonumber\\
&=\prod_{l=1}^{N_o}\psi^{(1,0)}(\textbf{s}_l)\prod_{m=1}^{N_x}\psi^{(0,1)}(\textbf{s}_{N_o+m}).\label{Eq_minimization}
\end{align}

For the square cavity, one obtains the coupled equations
\begin{align}
\psi^{(2,0)}=\frac{\rho^{o}(s_1,s_2)\rho^{o}(s_1,s_2+1)}{1-\rho^{x}(s_1,s_2)-\rho^{x}(s_1+1,s_2)+\psi^{(0,2)}},\nonumber \\
\psi^{(0,2)}=\frac{\rho^{x}(s_1,s_2)\rho^{x}(s_1+1,s_2)}{1-\rho^{o}(s_1,s_2)-\rho^{o}(s_1,s_2+1)+\psi^{(2,0)}}, \label{Eq_coupled}
\end{align}
with the solutions
\begin{align}
  	&\psi^{(0,2)}_{\pm}\left(\rho^{o}(1,1),\rho^{o}(1,2),\rho^{x}(1,1),\rho^{x}(2,1)\right)\nonumber\\
  	&=\frac{1}{2(1-\rho^{o}(1,1)-\rho^{o}(1,2))}\Bigl[-\bigl(\left(1-\rho^{x}(1,1)-\rho^{x}(2,1)\right)\left(1-\rho^{o}(1,1)-\rho^{o}(1,2)\right)\nonumber\\
&+\rho^{o}(1,1)\rho^{o}(1,2)-\rho^{x}(1,1)\rho^{x}(2,1))\bigr)
\pm\Bigl(\bigl((1-\rho^{x}(1,1)-\rho^{x}(2,1))\nonumber\\
&\cdot (1-\rho^{o}(1,1)-\rho^{o}(1,2))+\rho^{o}(1,1)\rho^{o}(1,2)-\rho^{x}(1,1)\rho^{x}(2,1)\bigr)^2\nonumber \\
&+4(1-\rho^{o}(1,1)-\rho^{o}(1,2))(1-\rho^{x}(1,1)-\rho^{x}(2,1))\rho^{x}(1,1)\rho^{x}(2,1)\Bigr)^\frac{1}{2}\Bigr],\nonumber\\ 
&\psi^{(2,0)}_{\pm}\left(\rho^{o}(1,1),\rho^{o}(1,2),\rho^{x}(1,1),\rho^{x}(2,1)\right)=\psi^{(0,2)}_{\pm}\left(\rho^{x}(1,1),\rho^{x}(2,1),\rho^{o}(1,1),\rho^{o}(1,2)\right). \end{align} 
The solutions  $\psi^{(2,0)}_{-}$ and $\psi^{(0,2)}_{-}$ can be neglected since they are not positive, as required for probabilities.
The free energy for the square cavity $\mathcal{F}^{2\times2}_{\text{L}}$ is consequently obtained by inserting $\psi^{(2,0)}_{+}$ and $\psi^{(0,2)}_{+}$ 
into Eq.~\ref{Eq_Levy_square}. 
Subtracting the ideal gas free energy defined in Eq.~\ref{Eq_ideal_free_functional} then gives the excess free energy of the square cavity
\begin{align*}
\mathcal{F}_{\text{ex}}^{2\times 2}=\mathcal{F}^{2\times2}_{\text{L}}-\mathcal{F}_{\text{id}}^{2\times2}
=: \Phi^{2\times 2}\left(\rho^{o}(s_1,s_2),\rho^{o}(s_1,s_2+1),\rho^{x}(s_1,s_2),\rho^{x}(s_1+1,s_2)\right).
\end{align*}
This new function $\Phi^{2\times 2}$ is a similar building block for the free energy functional as is $\Phi^{0\text{D}}$, but the explicit expression is rather lengthy and includes the use of $\psi^{(0,2)}_+$ and $\psi^{(2,0)}_+$. 
It is noticeable that already $\psi^{(0,2)}_+$ and $\psi^{(2,0)}_+$ cannot be written using the FMT or any other simple weighted densities and this feature is transmitted to $\Phi^{2\times2}$. 
Analogously to the notation in the previous chapter, we now introduce
\begin{align}
\nquadrat (s_1,s_2)&=(\rho^{o}(s_1,s_2),\rho^{o}(s_1,s_2+1),\rho^{x}(s_1,s_2),\rho^{x}(s_1+1,s_2)),\nonumber\\
\nleins (s_1,s_2)&=(\rho^{o}(s_1,s_2),\rho^{x}(s_1,s_2)),\nonumber\\
\nlzwei (s_1,s_2)&=(\rho^{o}(s_1,s_2),\rho^{x}(s_1+1,s_2)),\nonumber\\
\nldrei (s_1,s_2)&=(\rho^{o}(s_1,s_2+1),\rho^{x}(s_1,s_2)),\nonumber\\
\nlvier (s_1,s_2)&=(\rho^{o}(s_1,s_2+1),\rho^{x}(s_1+1,s_2)).\label{Eq_cavities_II}
\end{align}
Similarly to the tuples defined in Eq.~\ref{Eq_cavities_I}, we drop the explicit notation of the evaluation point $(s_1,s_2)$ in the following.
If one of the $\rho^{o}$ and one of the $\rho^{x}$ are zero,
one can interpret the system as a 0D cavity.
This implies (and can be verified by using the definition of $\Phi^{2\times2}$) that
$$\Phi^{2\times2}(\rho^{o}(s_1,s_2),0,\rho^{x}(s_1,s_2),0)=\Phi^{0\text{D}}(\nleins)$$
and analogously for $\nlzwei$, $\nldrei$ and $\nlvier$.

In order to obtain the free energy functional for abitrary situations,
we use LC's iterative subtraction scheme as presented before. The requirement is now that the free energy functional gives the exact free energy of all 0D cavities and of the square cavity.
For the free energy density, we choose the ansatz
\begin{align}
f_1^{\text{ex}}=\Phi^{0\text{D}}\left(\nkreuz\right)-\Phi^{0\text{D}}\left(\nwaagrecht\right)-\Phi^{0\text{D}}\left(\nsenkrecht\right)+\Phi^{2\times2}\left(\nquadrat\right),
\end{align}
which is the sum of the LC free energy density (exact for 0D cavities) as well as the value of the excess free energy density of the square cavity.
The evaluation for a fixed square cavity marked by \quadrat\ yields
 \begin{align*}
\left.\sum_{(s_1,s_2)\in \Z} f_1^{\text{ex}}(s_1,s_2)\right|_{\quadrat}&=\Phi^{2\times2}\left(\quadrat\right)+\Phi^{0\text{D}}\left(\quadrateins\right)+\Phi^{0\text{D}}\left(\quadratzwei\right)\nonumber\\
&+\Phi^{0\text{D}}\left(\quadratdrei\right)+\Phi^{0\text{D}}\left(\quadratvier\right).
\end{align*} 
In order to obtain a correct excess free energy density, calculating the square cavity correctly, the terms beyond $\Phi^{2\times2}\left(\quadrat\right)$ have to be eliminated. 
Therefore, an improved guess can be constructed by subtracting the additional terms. The new ansatz for the excess free energy density is consequently given by
\begin{align*}
f_2^{\text{ex}}=&=f_1^{\text{ex}}-\Phi^{0\text{D}}\left(\nleins\right)-\Phi^{0\text{D}}\left(\nlzwei\right)-\Phi^{0\text{D}}\left(\nldrei\right)-\Phi^{0\text{D}}\left(\nlvier\right)\\
&=\Phi^{0\text{D}}\left(\nkreuz\right)-\Phi^{0\text{D}}\left(\nwaagrecht\right)-\Phi^{0\text{D}}\left(\nsenkrecht\right)+\Phi^{2\times2}\left(\nquadrat\right)\\
&-\Phi^{0\text{D}}\left(\nleins\right)-\Phi^{0\text{D}}\left(\nlzwei\right)-\Phi^{0\text{D}}\left(\nldrei\right)-\Phi^{0\text{D}}\left(\nlvier\right).
\end{align*}
Applying once again the new excess free energy yields
\begin{align*}
\left.\sum_{(s_1,s_2)\in \Z} f_2^{\text{ex}}(s_1,s_2)\right|_{\quadrat}&=\Phi^{2\times2}\left(\quadrat\right)-2\Phi^{0\text{D}}\left(\quadratfuenf\right)
-2\Phi^{0\text{D}}\left(\quadratsechs\right)\\
&-2\Phi^{0\text{D}}\left(\quadratsieben\right)-2\Phi^{0\text{D}}\left(\quadratacht\right).
\end{align*}
Finally adding $2\Phi^{0\text{D}}(\nwaagrecht)+2\Phi^{0\text{D}}(\nsenkrecht)$ to the free energy density eliminates the unwanted terms on the rhs. The excess free energy density
\begin{align}
f^{\text{ex}}&=f_2^{\text{ex}}+2\Phi^{0\text{D}}(\nwaagrecht)+2\Phi^{0\text{D}}(\nsenkrecht)\nonumber\\
&=\Phi^{0\text{D}}\left(\nkreuz\right)+\Phi^{0\text{D}}\left(\nwaagrecht\right)+\Phi^{0\text{D}}\left(\nsenkrecht\right)+\Phi^{2\times2}\left(\nquadrat\right)\nonumber\\
&-\Phi^{0\text{D}}\left(\nleins\right)-\Phi^{0\text{D}}\left(\nlzwei\right)-\Phi^{0\text{D}}\left(\nldrei\right)-\Phi^{0\text{D}}\left(\nlvier\right). \label{Eq_density_new}
\end{align}
correctly determines the free energy of the square cavity and (as can be shown by explicit calculation) the free energy of a maximal 0D cavity.
Thus, it is a generalization of the Lafuente-Cuesta free energy density.
\subsection{Maximal cavities}
\begin{figure}
\centering
\includegraphics{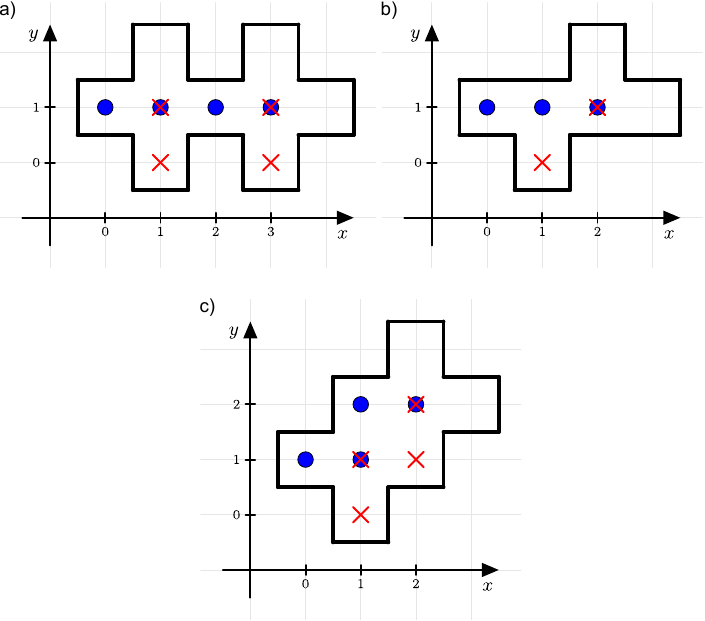}
\caption{All possible connected maximal cavities which can at most hold two particles at a time modulo symmetry operations as rotation and inversion.  \label{max_cavities}}
\end{figure} 

To our surprise, the new free energy density not only correctly determines the free energy of the 0D cavity and the square cavity, but also of all cavities 
in which at most two particles can be placed simultaneously.
To prove this, it is sufficient to consider only the maximal ones of these cavities, since all other cavities can be embedded in them. 
These maximal cavities are shown in Fig.~\ref{max_cavities}. 
In this section we limit ourselves to qualitative arguments for this phenomenon. 
The mathematical proof, which involves minimizing Levy's functional and comparing the result with the sum over the free energy density for each of the cavities, 
can be found in the supplement.

The reason for the correctness of free energy evaluated at the cavities shown in Fig.~\ref{max_cavities}.a is quite elaborate. 
Basically, the cavities consist of two maximal 0D cavities glued together. 
If both of them were isolated, it would not be surprising that their free energy is calculated correctly by the LC functional as well as by the new one.
The ``interaction'' between the two $0$D cavities is exclusively through horizontal particles placed at either $(1,1)$ or $(2,1)$.
Therefore, each configuration can be classified into one of the three cases, where a horizontal particle is found either at $(1,1)$, at $(2,1)$ or at neither. 
The set of configurations in which a horizontal particle is found at $(1,1)$ is disjoint from the set of configurations in which a horizontal particle is placed at $(2,1)$, since hard rods are not allowed to overlap. 
Eq.~\ref{Eq_constraint} implies that the sum of the probabilities of the configurations is either  $\rho^{o}(1,1)$ or respectively $\rho^{o}(2,1)$.
The normalization of the probabilities determines that the probability of finding no horizontal particle at both location is $1-\rho^{o}(1,1)-\rho^{o}(2,1)$.
The probabilities for configurations with a horizontal particle at location $(1,1)$ are thus only $\rho^{o}(1,1)$ times the probability of the other 0D cavity configuration, and a similar result holds for the configurations with a horizontal particle at location $(2,1)$.
If neither of the locations is occupied by a horizontal rod, the probabilities of the configurations can be treated as having two unconnected 0D cavities times $1-\rho^{o}(1,1)-\rho^{o}(2,1)$.
Therefore, the 1D-coupling is already correctly determined by the free energy functional of Lafuente and Cuesta.

The free energy of the cavity in Fig.~\ref{max_cavities}.b is described correctly already by the LC free energy functional. In short, 
the cavity is equivalent to a one-dimensional system (on the line $y=1$) of $L=2$ particles (the former horizontal rods) and $L=1$ particles (the former 
vertical rods), and the LC functional is exact in one dimension \cite{lafuente02}.

The last maximal cavities shown in Fig.~\ref{max_cavities}.c cannot be solved correctly by using the LC functional. This result is not surprising, since the square cavity can be embedded in it.
The cavity can be interpreted as two $0$D cavities and one square cavity which are interacting.
Both the free energy of the 0D cavity and the free energy of the square cavity by itself are correctly determined by the new free energy functional. 
It is interesting to note that no additional complexity seems to evolve by combining the two.

\section{Results}\label{sec4}
\begin{figure}[h]
\centering
\includegraphics[width=\textwidth]{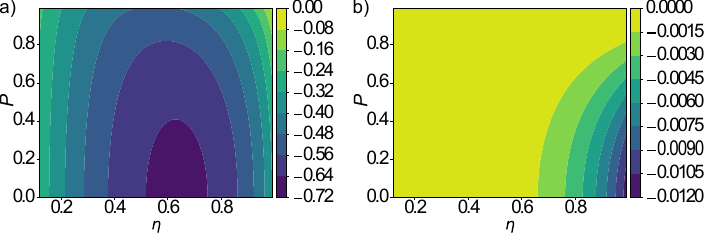}
\caption{The free energy density defined in Eq.~\ref{Eq_density_new} in a bulk system (a) and the difference of Lafuente and Cuesta's  free energy density and the one defined in Eq.~\ref{Eq_density_LC} in a bulk system (b) as function of the packing fraction $\eta$ and an order parameter $P$. \label{ABB_04}}
\end{figure}

To evaluate the performance of the new functional, we consider bulk systems and a particular confinement situation. 
In bulk,  $\rho^{o}(s_1,s_2)=\rho^{o}$ and $\rho^{x}(s_1,s_2)=\rho^{x}$ and we define the packing fraction to be $\eta=L(\rho^{o}+\rho^{x})$ and the order parameter $P=\frac{\rho^{x}-\rho^{o}}{\rho^{x}+\rho^{o}}$  as in \cite{oettel16}. It holds that $P \in [-1,1]$, where $P=1$ describes a system consisting only of vertical rods, $P=-1$ a system of horizontal rods and $P=0$ the isotropic system. 
Because of symmetry, it suffices to consider $P\in[0,1]$. The new free energy density is shown in Fig.~\ref{ABB_04}.a, and the difference between the LC and the new free energy density in Fig.~\ref{ABB_04}.b, both as a contourplot. Relative deviations are small, reaching about
4\% for $P=0$ (isotropic systems) at high densities. The difference is maximal at close packing ($\eta=1$), where one can compare to
exact value obtained by Fisher as a solution of the dimer problem \cite{fisher61}, $\lim_{\eta\to 1, \rho^{x}=\rho^{o}}\beta f^\text{exact}=-0.291$.
The LC free energy density in this limit is $-1.5\log(3)+2\log(2)\approx -0.262$ whereas the new free energy density is $1.5\log(3)-1.5\log(2)+\log(\sqrt{2}-1)\approx-0.273$.
It is closer to the exact value, but one may hypothesize that an expansion of the free energy density in $N_c$, the number of particles for a cavity for which the free energy is 
still exact, is not converging particularly fast beyond $N_c=1$, since the leading term, $N_c=1$, is given by the LC functional and accounts for  90\% of the
free energy density and the next term, $N_c=2$, as given by the new functional only accounts for about 38\% of the remaining difference to the exact free energy.

\begin{figure}[h]
\centering
\includegraphics{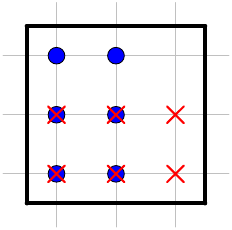}
\caption{Cavity of a $3\times 3$ box. \label{ABB_06}}
\end{figure}

\begin{figure}[h]
\centering
\includegraphics[width=\textwidth]{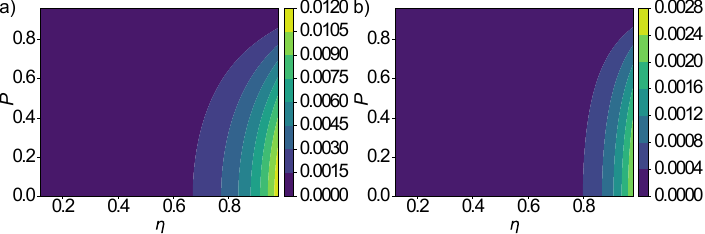}
\caption{The relative deviation between the exact value of the free energy and Lafuente and Cuesta's free energy (a) and the relative deviation between the exact value of the free energy and the free energy determined by using the new free energy functional (b) of a $3\times 3$ volume.\label{ABB_07}}
\end{figure}

A particular confinement situation is given by a cavity in the form of a $3\times 3$ box, see Fig.~\ref{ABB_06}. This cavity can hold up to 4 particles, and is treated exactly
neither by the LC nor the new functional. For a given density distribution inside the cavity, the exact free energy can be computed using 
Levy's functional defined in Eq.~\ref{Eq_Levy}
by a numerical minimization over all $\psi^{(N_o,N_x)}$ with $N_o+N_x\geq 2$ (in total 118 variables). We employed a standard gradient descent method and assumed a constant density $\rho_o$ of horizontal rods inside the cavity (at points marked by full circles in  Fig.~\ref{ABB_06}) and a constant density $\rho_x$ of vertical rods (at points marked by crosses in  Fig.~\ref{ABB_06}). 
In Fig.~\ref{ABB_07}.a resp. Fig.~\ref{ABB_07}.b, the relative deviation between the exact free energy and the LC free energy resp. the new free energy is shown as a contourplot in the $\eta$--$P$ plane. 
Again, the deviations are maximized for $P=0$ and $\eta \to 1$. For the LC functional, maximal deviations are around 1.2\%, for the new functional these are reduced to about 0.3\%. Thus, in contrast to the bulk system, a hypothetical expansion in $N_c$ seems to converge much better in this case: about 3/4 of the remaining difference between the exact and the LC free energy is accounted for by the new functional.

\section{Conclusion with remarks on related publications}\label{sec5}It is interesting to note that over the past decades until very recently, a variety of different approaches to the statistics of \text{bulk} systems of hard rods on lattices 
have been developed that all give the same free energy (entropy) per particle as fundamental measure theory with the LC functionals. Most notably, these include
a work by Guggenheim \cite{guggenheim44} valid for isotropic systems, a more complete treatment by Di Marzio \cite{dimarzio61} for unequal densities for the differently oriented rods
(anisotropic systems) and a solution on Bethe lattices, also for arbitrarily anisotropic systems \cite{dhar11}, see also the discussion in \cite{gschwind17}. These approaches are difficult to generalize to \text{inhomogeneous} situations which illustrates the advantage of density functional techniques. 
In order to compare the type of approximations in these statistical approaches to the one in the LC functionals (exact treatment of 0D cavities), it is instructive to briefly summarize one of these approaches.  

In 1961, Di Marzio determined an approximation for the probability of inserting a hard rod into a hard rod system on a 2D cubic lattice (i.e., a subvolume of $\Z^2$), as well as on a 3D lattice \cite{dimarzio61}. 
For this calculation, essentially he approximates that rods orthogonal to the inserted rod can be split into $L$ rods of size 1, randomly distributed on the lattice.
This corresponds to treating the orthogonal rods as rods of length $1$ with a bulk density of $L$ times the original bulk density.
The insertion probability for a rod of orientation $k$ is determined by first calculating the probability of having an empty site at a random location within the lattice. 
Afterwards, this quantity is multiplied by the conditional probability of having an empty site if the neighboring site of orientation $k$ is empty.
Repeating this procedure  $(L-1)$ times in a bulk system on a 2D lattice yields the probability
\begin{align*}
p_k=\frac{(1-L\rho^{o}-L\rho^{x})^{L}}{(1-(L-1)\rho^{k})^{L-1}}
\end{align*}
for inserting a rod with orientation $k$.
Using the potential distribution theorem (see for example \cite{hansen13}), this quantity is related to the excess chemical potential by
\begin{align*}
\mu^{\text{ex}}_k=\log(p_k)=L\log(1-L(\rho^{x}+\rho^{o}))-(L-1)\log(1-(L-1)\rho^{k}).
\end{align*}
and thus the chemical potential by
\begin{align*}
\mu_k=\mu_k^{\text{id}}+\mu_k^{\text{ex}}=\rho^{k}\log\left(\rho^{k}\right)+L\log\left(1-L(\rho^{x}+\rho^{o})\right)-(L-1)\log\left(1-(L-1)\rho^{k}\right),
\end{align*}
and this is identical to the bulk excess chemical potential from the LC functional. Admittedly, the connection between the Di Marzio and the LC approximation is unclear at this point.

\begin{figure}
\centering
\includegraphics{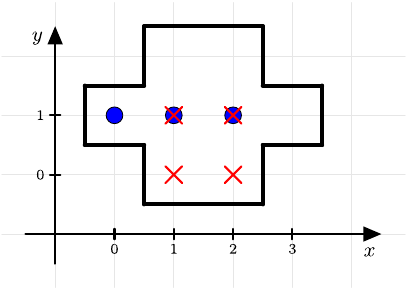}
\caption{Cavity considered to determine the exclusion spectrum function for dimers by Riccardo et al. \cite{riccardo23}.\label{ABB_08}}
\end{figure}

In more recent years, the Di Marzio-LC form of the bulk entropy has appeared in several comparisons to simulation results of the Ramirez-Pastor group,
see e.g. \cite{longone12,pasinetti23,riccardo23}. We note that in the very recent work of Riccardo et al. \cite{riccardo23} it is yet another guise that the Di Marzio-LC form appears.
The authors aimed at estimating the exclusion spectrum function, which is defined to be the average number of excluded states per particle, depending on packing fraction. 
It is furthermore shown, that this quantity is bijectively related to the chemical potential. Their method is based on the determination of configuration probabilities 
in the cavity shown in Fig.~\ref{ABB_08} for $L=2$. As indicated by the circles and crosses in the cavity, it is assumed that neither at $(1,0)$ nor at $(1,2)$ 
a horizontal dimer is placed or equivalently only configurations in which neither of the two rods appears are considered.
By ignoring these configurations, Riccardo et al. end up with a mean field approach similar to Di Marzio's, since they do not consider the indirect interaction between 
horizontal rods that do not share the same value on the $y$-axis. Thus, the vertical rods can again be interpreted as rods of length $L=1$ with a density twice that of 
the vertical rods in the original system. Consequently, it is not surprising that the authors end up with the same chemical potential as Di Marzio. 
In fact, if one applies Levy's functional to the cavity and includes only the same configurations as in \cite{riccardo23}, one will see that the free energy functional 
coincides with that of Lafuente an Cuesta. This perhaps better shows that the considerations in the bulk all share the same structure and the same chemical potential.

These digressions bring us to the main summary of the present paper.
By using Levy's functional we have constructed an extension of the Lafuente and Cuesta fundamental measure theory. Using that for bulk systems, it was finally possible to 
improve Di Marzio's approximations, but, as a density functional, it is applicable to non-bulk systems. On the other hand, the shown method of solving Levy's functional 
for bounded volumes and then using Lafuente and Cuesta's method to find an excess free energy density can, in principle, be extended to larger cavities. 
However, it has proven difficult to find an analytical solution for the probabilities of the configurations inside cavities beyond those cavities that can hold two 
particles simultaneously. 
Since the number of configurations grows almost exponentially with the volume of a cavity, there seems to be a natural limit to finding exact free energy functionals with the method shown here. The minimization of Levy's functional defines a new equation for every configuration (excluding configurations with less than two particles). If it is possible to combine these equations into one polynomial depending on only one variable, as is the case for all cavities in this work, the task of finding an analytical solution of the free energy for a given one particle density distribution reduces to a question of reducibility and solvability of this polynomial. Both are well-discussed algebraic problems and at least the solvability can be determined using Galois theory.

As mentioned in the derivation of the new free energy functional, similar assumptions can be made to generate a free energy functional for larger rod sizes by 
considering two-particle cavities. Larger rod sizes are thermodynamically more interesting, since Monte Carlo simulations have established that there is a phase transition to a nematic phase (demixed pahse between horizontal and vertical rods) for rod lengths $L\geq 7$ \cite{ghosh07,kundu13}. 
However, the free energy functional of Lafuente and Cuesta predicts a phase transition for $L\geq 4$ \cite{oettel16}. Preliminary calculations show that the spinodal point is shifted to higher packing fractions for the new free energy functionals compared to the functionals of Lafuente and Cuesta. Nevertheless, it appears for all $L\geq 4$. 
It is a plausible assumption that a free energy functional that takes into account square cavities of the length of the respective rod size may fix this behavior.
This is, however, beyond the scope of the present paper.

\section*{Data Avalaibility Statement}
The numerical data in this study are available from the authors upon reasonable request.
\section*{Acknowledgements}
We  acknowledge  support  by  the  Deutsche  Forschungsgemeinschaft  (DFG, Grant  No.Oe 285/6-1  within  SPP  2171). 
\section*{Conflict of Interest Statement}
The authors declare no conflicts of interest.
\bibliography{Bib}

\begin{appendices}
\section{Supplement}
In this chapter it is shown that the free energy functional of the eight maximal cavities, which can hold two particles at the same time, is already correctly calculated by the energy density defined in Eq.~\ref{Eq_density_new}. 
The eight cavities are obtained by using symmetry operations as rotations and inversions on the cavities in Fig.~\ref{max_cavities}.
To proceed, we calculate the free energy functional using Levy's formula given in Eq.~\ref{Eq_Levy}. To verify the correctness of the new free energy functional, one could use brute force and insert the presented configuration probabilities into Levy's formula. Nevertheless, we want to discuss the probabilities of the configurations in more detail, so that the reader may develop an intuition for them. The position of the cavities is arbitrary. For the sake of simplicity, we will place them in $\Z^2$, as shown in Fig.~\ref{max_cavities}. 
For the first cavity, shown in Fig.~\ref{max_cavities}.a, the constraint $\psi \to \rho$  in Levy's formula yields
\begin{align*}
\rho^{o}(0,1)&=\psi^{(1,0)}(0,1)+\psi^{(2,0)}(0,1;2,1)+\psi^{(2,0)}(0,1;3,1)+\psi^{(1,1)}(0,1;3,0)\\
&+\psi^{(1,1)}(0,1;3,1),\\
\rho^{o}(1,1)&=\psi^{(1,0)}(1,1)+\psi^{(2,0)}(1,1;3,1)+\psi^{(1,1)}(1,1;3,0)+\psi^{(1,1)}(1,1;3,1),\\
\rho^{o}(2,1)&=\psi^{(1,0)}(2,1)+\psi^{(2,0)}(0,1;2,1)+\psi^{(1,1)}(2,1;1,0)+\psi^{(1,1)}(2,1;1,1),\\
\rho^{o}(3,1)&=\psi^{(1,0)}(3,1)+\psi^{(2,0)}(0,1;3,1)+\psi^{(2,0)}(1,1;3,1)+\psi^{(1,1)}(3,1;1,0),\\
&+\psi^{(1,1)}(3,1;1,1),\\
\rho^{x}(0,1)&=\psi^{(0,1)}(1,0)+\psi^{(0,2)}(1,0;3,0)+\psi^{(0,2)}(1,0;3,1)+\psi^{(1,1)}(2,1;1,0)\\
&+\psi^{(1,1)}(3,1;0,1),\\
\rho^{x}(1,1)&=\psi^{(0,1)}(1,1)+\psi^{(0,2)}(1,1;3,0)+\psi^{(0,2)}(1,1;3,1)+\psi^{(1,1)}(2,1;1,1)\\
&+\psi^{(1,1)}(3,1;1,1),\\
\rho^{x}(3,0)&=\psi^{(0,1)}(3,0)+\psi^{(0,2)}(1,0;3,0)+\psi^{(0,2)}(1,1;3,0)+\psi^{(1,1)}(0,1;3,0)\\
&+\psi^{(1,1)}(1,1;3,0),\\
\rho^{x}(3,1)&=\psi^{(0,1)}(3,1)+\psi^{(0,2)}(1,0;3,1)+\psi^{(0,2)}(1,1;3,1)+\psi^{(1,1)}(0,1;3,1)\\
&+\psi^{(1,1)}(1,1;3,1).
\end{align*}
For simplicty of notation we introduce $\textbf{M}_{N_o+N_x}=(\textbf{s}_1;\dots;\textbf{s}_{N_o};\textbf{s}_{N_o+1};\dots;\textbf{s}_{N_o+N_x})$, $\textbf{M}_{N_o}=(\textbf{s}_1;\dots;\textbf{s}_{N_o})$ and $\textbf{M}_{N_x}=(\textbf{s}_{N_o+1};\dots;\textbf{s}_{N_o+N_x})$. By using that $\psi^{(0,0)}=1-\sum_{\psi\neq\psi^{(0,0)}}\psi$, the minimization in Levy's formula implies as already stated in Eq.~\ref{Eq_minimization} the following dependencies
\begin{align*}
&\psi^{(N_o,N_x)}(\textbf{M}_{N_o+N_x})\left(\psi^{(0,0)}\right)^{(N_o+N_x-1)}=\prod_{l=1}^{N_o}\psi^{(1,0)}(\textbf{s}_l)\prod_{m=1}^{N_x}\psi^{(0,1)}(\textbf{s}_{N_o+m})
\end{align*}
for all two-particle configurations appearing in the constraint above. The solutions for these are of the type
\begin{align}
&\psi^{(N_o,N_x)}(\textbf{M}_{N_o+N_x})=\prod_{l=1}^{N_o}\rho^{o}(\textbf{s}_l)\prod_{m=1}^{N_x}\rho^{x}(\textbf{s}_{N_o+m})\nonumber\\
&\cdot \left(\prod_{\substack{(s_1,s_2)\in \Z,\\ (s_1,s_2),(s_1+1,s_2)\notin \textbf{M}_{N_o},\\(s_1+1,s_2),(s_1+1,s_2-1)\notin\textbf{M}_{N_x}}}\left(1-\sum_{\rho^{i}(t_1,t_2)\in\nkreuz}\rho^{i}(t_1,t_2)\right)\right)\nonumber\\
&\cdot\left( \prod_{\substack{\textbf{s}\in \Z,\\ \textbf{s} \notin \textbf{M}_{N_o}}}\left(1-\sum_{\rho^{i}(t_1,t_2)\in\nwaagrecht}\rho^{i}(t_1,t_2)\right) \prod_{\substack{\textbf{s}\in \Z,\\ \textbf{s} \notin \textbf{M}_{N_x}}}\left(1-\sum_{\rho^{i}(t_1,t_2)\in\nsenkrecht}\rho^{i}(t_1,t_2)\right) \right)^{-1},\label{Eq_probs}
\end{align}
where we set $\rho^{o}(\textbf{s})=0$ if no horizontal rod can be placed within the cavity at $\textbf{s}$ and similarly for $\rho^{x}(\textbf{s})$. 
Consequently, using the functional equation of the logarithm, Levy's formula yields
\begin{align*}
&\mathcal{F}_{\text{L}}[\rho]=\sum_{N_o+N_x\geq0}\sum_{\textbf{M}_{N_o+N_x}\in C^{N_o,N_x}}S_\text{n}(\psi^{(N_o,N_x)}(\textbf{M}_{N_o+N_x}))\\
&=\sum_{N_o+N_x\geq0}\sum_{\textbf{M}_{N_o+N_x}\in C^{N_o,N_x}}\psi^{(N_o,N_x)}(\textbf{M}_{N_o+N_x})\\
&\cdot\left(\sum_{k=1}^{N_o}\log(\rho^{o}(\textbf{s}_k))+\sum_{m=1}^{N_x}\log(\rho^{x}(\textbf{s}_{N_o+m})\right.\\
&+\sum_{\substack{(s_1,s_2)\in \Z,\\ (s_1,s_2),(s_1+1,s_2)\notin \textbf{M}_{N_o},\\(s_1+1,s_2),(s_1+1,s_2-1)\notin \textbf{M}_{N_x}}}\log\left(1-\sum_{\rho^{i}(t_1,t_2)\in\nkreuz}\rho^{i}(t_1,t_2)\right)\\
&\left.-\sum_{\substack{\textbf{s}\in \Z,\\ \textbf{s} \notin \textbf{M}_{N_o}}}\log\left(1-\sum_{\rho^{i}(t_1,t_2)\in\nwaagrecht}\rho^{i}(t_1,t_2)\right)-\sum_{\substack{\textbf{s}\in \Z,\\ \textbf{s} \notin\textbf{M}_{N_x}}}\log\left(1-\sum_{\rho^{i}(t_1,t_2)\in\nsenkrecht}\rho^{i}(t_1,t_2)\right)\right),
\end{align*}
where $C^{N_o,N_x}$ is the set of all possible configurations with $N_o$ horizontal and $N_x$ vertical rods.
For a given $\textbf{s}$ inside the cavity, the term $\log(\rho^{o}(\textbf{s}))$ appears with the coefficients  $\psi^{(N_o,N_x)}(\textbf{M}_{N_o+N_x})$
for $N_o\geq 1$ and $N_x\geq0$ if $\textbf{M}_{N_o+N_x}\in C_{o,\textbf{s}}^{N_o,N_x}$.
By definition of the constraint, one obtains, that this sum yields $\rho^{o}(\textbf{s})$.
A similar argument can be made for the term $\log(\rho^{x}(\textbf{s}))$. Analogously, the term $\log(1-\rho^{o}(\textbf{s}))$ appears in the sum with the probabilities $\psi^{(N_o,N_x)}(\textbf{M}_{N_o+N_x})$, if $\textbf{M}_{N_o+N_x}\notin C_{o,\textbf{s}}^{N_o,N_x}$. Consequently, by using the normalization of the probability, the sum over these configurations is just $1-\rho^{o}(\textbf{s})$, and similarly the term $\log(1-\rho^{x}(\textbf{s}))$ appears with the coefficient $1-\rho^{x}(\textbf{s})$. Finally, the term $\log(1-\rho^{o}(s_1,s_2)-\rho^{o}(s_1+1,s_2)-\rho^{x}(s_1+1,s_2)-\rho^{x}(s_1+1,s_2-1))$ appears with coefficients  $\psi^{(N_o,N_x)}(\textbf{M}_{N_o+N_x})$ if $\textbf{M}_{N_o+N_x}\notin C_{o,(s_1,s_2)}^{N_o,N_x}\cup C_{o,(s_1+1,s_2)}^{N_o,N_x}\cup C_{x,(s_1+1,s_2)}^{N_o,N_x}\cup C_{x,(s_1+1,s_2-1)}^{N_o,N_x}$. Since all four sets mentioned above are pairwise disjoint due to the hard rod condition, the sum over all these coefficients yields $(1-\rho^{o}(s_1,s_2)-\rho^{o}(s_1+1,s_2)-\rho^{x}(s_1+1,s_2)-\rho^{x}(s_1+1,s_2-1))$. This implies that the free energy functional has the form
\begin{align*}
\mathcal{F}[\rho]&=\sum_{\textbf{s}\in \Z^2}  \rho^{o}(\textbf{s})\log(\rho^{o}(\textbf{s}))+\rho^{o}(\textbf{s})+\rho^{x}(\textbf{s})\log(\rho^{x}(\textbf{s}))+\rho^{x}(\textbf{s})\\
&+\sum_{(s_1,s_2)\in \Z^2} \Phi^{0\text{D}}\left(\nkreuz\right)-\Phi^{0\text{D}}\left(\nwaagrecht\right)-\Phi^{0\text{D}}\left(\nsenkrecht\right)\\
&=\sum_{\textbf{s}\in \Z} f^{\text{id}}(\textbf{s})+f^{\text{ex}}(\textbf{s}).
\end{align*}
Consequently,  the free energy is exactly calculated by the functional of Lafuente and Cuesta. As an extension of their functional, this holds also for the functional introduced here. The calculation can be easily transferred to the rotation of the cavity.

The second maximal cavity is shown in Fig.~\ref{max_cavities}.b. The constraint 
implies that 
\begin{align*}
\rho^{o}(0,1)&=\psi^{(1,0)}(0,1)+\psi^{(2,0)}(0,1;2,1)+\psi^{(1,1)}(0,1;2,1),\\
\rho^{o}(1,1)&=\psi^{(1,0)}(1,1),\\
\rho^{o}(2,1)&=\psi^{(1,0)}(2,1)+\psi^{(2,0)}(0,1;2,1)+\psi^{(1,1)}(2,1;1,0),\\
\rho^{x}(1,0)&=\psi^{(0,1)}(1,0)+\psi^{(0,2)}(1,0;2,1)+\psi^{(1,1)}(2,1;1,0),\\
\rho^{x}(2,1)&=\psi^{(0,1)}(2,1)+\psi^{(0,2)}(1,0;2,1)+\psi^{(1,1)}(0,1;2,1).
\end{align*}
The probabilities obtained by minimizing Levy's functional have the same form as described in Eq.~\ref{Eq_probs}. 
Consequently, the argument, why the functional is correctly determined by Lafuente and Cuesta's as well as our description is analogous to that of the previous cavity. The same holds for all symmetry operations used on the cavity.

For the third cavity, shown in Fig.~\ref{max_cavities}.c , we repeat the calculations which were made for the first and second cavity. The constraint implies
\begin{align*}
\rho^{o}(0,1)&=\psi^{(1,0)}(0,1)+\psi^{(2,0)}(0,1;1,2)+\psi^{(2,0)}(0,1;2,2)+\psi^{(1,1)}(0,1;2,1)\\
&+\psi^{(1,1)}(0,1;2,2),\\
\rho^{o}(1,1)&=\psi^{(1,0)}(1,1)+\psi^{(2,0)}(1,1;1,2)+\psi^{(2,0)}(1,1;2,2)+\psi^{(1,1)}(1,1;2,2),\\
\rho^{o}(1,2)&=\psi^{(1,0)}(1,2)+\psi^{(2,0)}(0,1;1,2)+\psi^{(2,0)}(1,1;1,2)+\psi^{(1,1)}(1,2;1,0),\\
\rho^{o}(2,2)&=\psi^{(1,0)}(2,2)+\psi^{(2,0)}(0,1;2,2)+\psi^{(2,0)}(1,1;2,2)+\psi^{(1,1)}(2,2;1,0)\\ 
&+\psi^{(1,1)}(2,2;1,1),\\
\rho^{x}(1,0)&=\psi^{(0,1)}(1,0)+\psi^{(0,2)}(1,0;2,1)+\psi^{(0,2)}(1,0;2,2)+\psi^{(1,1)}(1,2;1,0)\\ 
&+\psi^{(1,1)}(2,2;1,0),\\
\rho^{x}(1,1)&=\psi^{(0,1)}(1,1)+\psi^{(0,2)}(1,1;2,1)+\psi^{(0,2)}(1,1;2,2)+\psi^{(1,1)}(2,2;1,1),\\
\rho^{x}(2,1)&=\psi^{(0,1)}(2,1)+\psi^{(0,2)}(1,0;2,1)+\psi^{(0,2)}(1,1;2,1)+\psi^{(1,1)}(0,1;2,1),\\
\rho^{x}(2,2)&=\psi^{(0,1)}(2,2)+\psi^{(0,2)}(1,0;2,2)+\psi^{(0,2)}(1,1;2,2)+\psi^{(1,1)}(0,1;2,2)\\ 
&+\psi^{(1,1)}(1,1;2,2).
\end{align*}
Again, we find that the minimization after the free parameters (i.e., all probabilities for those configurations where two particles are located in the cavity) satisfies the following requirements on the probabilities
\begin{align*}
&\psi^{(N_o,N_x)}(\textbf{M}_{N_o+N_x})\left(\psi^{(0,0)}\right)^{(N_o+N_x-1)}=\prod_{l=1}^{N_o}\psi^{(1,0)}(\textbf{s}_l)\prod_{m=1}^{N_x}\psi^{(0,1)}(\textbf{s}_{N_o+m}).
\end{align*}
Similar to Eq.~\ref{Eq_coupled}, this yields the following coupled equations
\begin{align*}
\psi^{(2,0)}(1,1;1,2)&=\frac{\rho^{o}(1,1)\rho^{o}(1,2)}{1-\rho^{x}(1,1)-\rho^{x}(2,1)+\psi^{(0,2)}(1,1;2,1)},\\
\psi^{(0,2)}(1,1;2,1)&=\frac{\rho^{x}(1,1)\rho^{x}(2,1)}{1-\rho^{o}(1,1)-\rho^{o}(1,2)+\psi^{(2,0)}(1,1;1,2)}
\end{align*}
and thus, we get $\psi^{(2,0)}(1,1;1,2)=\psi^{2\times 2}(\rho^{o}(1,1),\rho^{o}(1,2),\rho^{x}(1,1)\rho^{x}(2,1))$ and  $\psi^{(0,2)}(1,1;2,1)=\psi^{2\times 2}(\rho^{x}(1,1)\rho^{x}(2,1),\rho^{o}(1,1),\rho^{o}(1,2))$. In contrast to the other cavities discussed in this section, 
a description of the configuration probabilities as done for the previous two types of maximal cavities is no longer possible, since 
neither $\psi^{(2,0)}(1,1;1,2)$ nor $\psi^{(0,2)}(1,1;2,1)$ has the suitable form. Nevertheless, restricted to this cavity, structures can be found in the description 
of the configurations' probabilities as well. For example, the probability of not finding a particle in the cavity is given by 
\begin{align*}
\psi^{(0,0)}&=\prod_{(s_1,s_2)\in\Z^2}\left(1-\sum_{\rho^{i}(t_1,t_2)\in\nkreuz}\rho^{i}(t_1,t_2)\right)\left(1-\sum_{\rho^{i}(t_1,t_2)\in\nwaagrecht}\rho^{i}(t_1,t_2)\right)\\
&\cdot\left(1-\sum_{\rho^{i}(t_1,t_2)\in\nsenkrecht}\rho^{i}(t_1,t_2)\right)
\\&\cdot\left(1-\sum_{\rho^{i}(t_1,t_2)\in\nquadrat}\rho^{i}(t_1,t_2)+\psi^{(2,0)}(\nquadrat)+\psi^{(0,2)}(\nquadrat)\right)\\
&\cdot\left[\left(1-\sum_{\rho^{i}(t_1,t_2)\in\nleins}\rho^{i}(t_1,t_2)\right)\left(1-\sum_{\rho^{i}(t_1,t_2)\in\nlzwei}\rho^{i}(t_1,t_2)\right)\right.\\
&\cdot\left.\left(1-\sum_{\rho^{i}(t_1,t_2)\in\nldrei}\rho^{i}(t_1,t_2)\right)\left(1-\sum_{\rho^{i}(t_1,t_2)\in\nlvier}\rho^{i}(t_1,t_2)\right)\right]^{-1},
\end{align*}
where we used the definitions in Eqs.~\ref{Eq_cavities_I} and \ref{Eq_cavities_II} for tuples of one-particle densities. In particular, each of the the mentioned tuples of one-particle densities depends on $(s_1,s_2)$ as it is stated there. For the description of furhter probabilities ofconfigurations, we introduce the shift operator $T_{(p,q)}$, which shifts the position of each of the one-particle densities in the tuples in the horiziontal direction by $p$ steps and in the vertical direction by $q$ steps. For example, one obtains
\begin{align*}
T_{(p,q)}\nlzwei&=T_{(p,q)}(\rho^{o}(s_1,s_2),\rho^{x}(s_1+1,s_2))\\
&=(\rho^{o}(s_1+p,s_2+q),\rho^{x}(s_1+p+1,s_2+q)).
\end{align*}
If there is no rod at any of the locations $(1,1)$, $(1,2)$, $(2,1)$, the probability of the configuration can be described as
\begin{align*}
&\psi^{(N_o,N_x)}(\textbf{M}_{N_o+N_x})=
\left(\psi^{(0,0)}\prod_{(s_1,s_2)\in\textbf{M}_{N_o}}\rho^{o}(s_1,s_2)\left(1-\sum_{\rho^{i}(t_1,t_2)\in\nleins}\rho^{i}(t_1,t_2)\right)\right.\\ 
&\left(1-\sum_{\rho^{i}(t_1,t_2)\in\nlzwei}\rho^{i}(t_1,t_2)\right)\left(1-\sum_{\rho^{i}(t_1,t_2)\in T_{(0,-1)}\nldrei}\rho^{i}(t_1,t_2)\right)\\
&\left.\left(1-\sum_{\rho^{i}(t_1,t_2)\in T_{(0,-1)}\nlvier}\rho^{i}(t_1,t_2)\right)\right)\left(\prod_{(s_1,s_2)\in\textbf{M}_{N_x}}\rho^{x}(s_1,s_2)\right.\\
&\left.\prod_{(\rho^{o}(t_1,t_2),\rho^{x}(s_1,s_2))\in \{\nleins,\nlzwei,\nldrei,\nlvier\}} ((1-\rho^{o}(t_1,t_2)-\rho^{x}(s_1,s_2))\right)\\
&\cdot\left[\prod_{(s_1,s_2)\in\textbf{M}_{N_o}}(1-\rho^{o}(s_1,s_2))\left(1-\sum_{\rho^{i}(t_1,t_2)\in T_{(0,-1)}\nkreuz}\rho^{i}(t_1,t_2)\right)\right.\\
&\left.\left(1-\sum_{\rho^{i}(t_1,t_2)\in T_{(-1,-1)}\nkreuz}\rho^{i}(t_1,t_2)\right)\right]^{-1}\\
& \cdot\left[\prod_{(s_1,s_2)\in\textbf{M}_{N_x}}(1-\rho^{x}(s_1,s_2))\left(1-\sum_{\rho^{i}(t_1,t_2)\in T_{(-1,0)}\nkreuz}\rho^{i}(t_1,t_2)\right)\right.\\
&\left.\left(1-\sum_{\rho^{i}(t_1,t_2)\in T_{(-1,-1)}\nkreuz}\rho^{i}(t_1,t_2)\right)\right]^{-1}.
\end{align*}
If one of the horizontal rods is located at $\textbf{p}\in\{(1,1),(1,2)\}$, the probability of the configuration is given by
\begin{align*}
&\psi^{(N_o,N_x)} \left(\textbf{s}_1;\dots;\textbf{p};\dots;\textbf{s}_{N_o};\textbf{s}_{N_o+1};\dots;\textbf{s}_{N_o+N_x}\right)\\
&=\psi^{(N_o-1,N_x)}\left(\textbf{s}_1;\dots;\textbf{s}_{N_o};\textbf{s}_{N_o+1};\dots;\textbf{s}_{N_o+N_x}\right)\left(\rho^{o}(\textbf{p})-\psi^{(2,0)}\left(T_{(-s_1+1,-s_2+1)}\nquadrat\right)\right)\nonumber\\
&\cdot \Biggl[1-\sum_{\rho^{i}(t_1,t_2)\in T_{(-s_1+1,s_2+1)}\nquadrat}\rho^{i}(t_1,t_2)+\psi^{(2,0)}\left(T_{(-s_1+1,-s_2+1)}\nquadrat\right)\nonumber\\&
+\psi^{(2,0)}\left(T_{(-s_1+1,-s_2+1)}\nquadrat\right)\Biggr]^{-1}
\end{align*}
and analogously if one vertical rod is located at $\textbf{p}\in\{(1,1),(2,1)\}$, the configuration's probability is
\begin{align*}
&\psi^{(N_o,N_x)}(\textbf{s}_1;\dots;\textbf{s}_{N_o};\textbf{s}_{N_o+1};\dots;\textbf{p};\dots;\textbf{s}_{N_o+N_x})\\
&=\psi^{(N_o,N_x-1)}(\textbf{s}_1;\dots;\textbf{s}_{N_o};\textbf{s}_{N_o+1};\dots;\textbf{s}_{N_o+N_x})\cdot\left(\rho^{x}(\textbf{p})-\psi^{(0,2)}\left(T_{(-s_1+1,-s_2+1)}\nquadrat\right)\right)\\
&\cdot\Biggl[1-\sum_{\rho^{i}(t_1,t_2)\in T_{(-s_1+1,s_2+1)}\nquadrat}\rho^{i}(t_1,t_2)+\psi^{(2,0)}\left(T_{(-s_1+1,-s_2+1)}\nquadrat\right)\\
&+\psi^{(2,0)}\left(T_{(-s_1+1,-s_2+1)}\nquadrat\right)\Biggr]^{-1}.
\end{align*}
Applying once again the functional equation of the logarithm on the various probabilities allows a reordering of the sum in Levy's functional. Similar arguments as for the previous cavities can be made in order to sum the probabilities of the configurations at least for some of the terms occurring in Levy's functional.

Therefore, the exact energy functional for the given cavity is
\begin{align*}
&\beta \mathcal{F}[\rho]=\sum_\psi S_\text{n}(\psi)\\
&=\left(\psi^{(0,0)}+\psi^{(1,0)}(1,2)+\psi^{(1,0)}(2,2)+\psi^{(0,1)}(2,1)+\psi^{(0,1)}(2,2)\right)\\
&\cdot\log\left(1-\rho^{o}(0,1)-\rho^{o}(1,1)-\rho^{x}(1,0)-\rho^{x}(1,1)\right)\\
&+\left(\psi^{(0,0)}+\psi^{(1,0)}(0,1)+\psi^{(1,0)}(1,1)+\psi^{(0,1)}(1,0)+\psi^{(0,1)}(1,1)\right)\\
&\cdot \log(1-\rho^{o}(1,2)-\rho^{o}(2,2)-\rho^{x}(2,1)-\rho^{x}(2,2))\\
&+\left(\psi^{(0,0)}+\psi^{(1,0)}(0,1)+\psi^{(1,0)}(2,2)+\psi^{(0,1)}(1,0)+\psi^{(0,1)}(2,2)\right.\\
&\left.+\psi^{(2,0)}(0,1;2,2)+\psi^{(0,2)}(1,0;2,2)+\psi^{(1,1)}(0,1;2,2)+\psi^{(1,1)}(2,2;1,1)\right)\\
&\cdot \log\Bigl(1-\rho^{o}(1,1)-\rho^{(1,0)}(1,2)-\rho^{x}(1,1)-\rho^{x}(2,1)+\psi^{(2,0)}(1,1;1,2)\\
&+\psi^{(0,2)}(1,1;2,1) \Bigr)-\Bigl(\psi^{(0,0)}+\psi^{(1,0)}(0,1)+\psi^{(1,0)}(1,2)+\psi^{(1,0)}(2,2)+\psi^{(0,1)}(1,0)\\
&+\psi^{(0,1)}(2,1)+\psi^{(0,1)}(2,2)+\psi^{(2,0)}(0,1;1,2)+\psi^{(2,0)}(0,1;2,2)+\psi^{(0,2)}(1,0;2,1)\\
&+\psi^{(0,2)}(1,0;2,2)+\psi^{(1,1)}(0,1;2,1)+\psi^{(1,1)}(0,1;2,2)+\psi^{(1,1)}(1,2;1,0)\\
&+\psi^{(1,1)}(2,2;1,0)\Bigr)\cdot\log(1-\rho^{o}(1,1)-\rho^{x}(1,1))-\Bigl(\psi^{(0,0)}+\psi^{(1,0)}(0,1)+\psi^{(1,0)}(1,1)\\
&+\psi^{(1,0)}(2,2)+\psi^{(0,1)}(1,0)+\psi^{(0,1)}(2,1)+\psi^{(0,1)}(2,2)+\psi^{(2,0)}(0,1;2,2)\\
&+\psi^{(2,0)}(1,1;2,2)+\psi^{(0,2)}(1,0;2,2)+\psi^{(0,2)}(1,1;2,2)+\psi^{(1,1)}(0,1;2,2)\\
&+\psi^{(1,1)}(1,1;2,2)+\psi^{(1,1)}(2,2;1,0)+\psi^{(1,1)}(2,2;1,1)\Bigr)\log(1-\rho^{o}(1,2)-\rho^{x}(2,1))\\
&+\psi^{(2,0)}(1,1;1,2)\log(\psi^{(2,0)}(1,1;1,2))+\psi^{(0,2)}(1,1;2,1)\log\left(\psi^{(0,2)}(1,1;2,1)\right)\\
&+\left(\psi^{(1,0)}(1,1)+\psi^{(2,0)}(1,1;2,2)+\psi^{(1,1)}(1,1;2,2)\right)\log\left(\rho^{o}(1,1)-\psi^{(2,0)}(1,1;1,2)\right)\\
&+\left(\psi^{(1,0)}(1,2)+\psi^{(2,0)}(0,1;1,2)+\psi^{(1,1)}(1,1;2,2)\right)\log\left(\rho^{o}(1,2)-\psi^{(2,0)}(1,1;1,2)\right)\\
&+\left(\psi^{(0,1)}(1,1)+\psi^{(0,2)}(1,1;2,2)+\psi^{(1,1)}(2,2;1,1)\right)\log\left(\rho^{x}(1,1)-\psi^{(0,2)}(1,1;2,1)\right)\\
&+\left(\psi^{(0,1)}(2,1)+\psi^{(0,2)}(1,0;2,1)+\psi^{(1,1)}(0,1;2,1)\right)\log\left(\rho^{x}(2,1)-\psi^{(0,2)}(1,1;2,1)\right)\\
&+\Bigl(\psi^{(1,0)}(0,1)+\psi^{(2,0)}(0,1;1,2)+\psi^{(2,0)}(0,1;2,2)+\psi^{(1,1)}(0,1;2,1)\\
&+\psi^{(1,1)}(0,1;2,2)\Bigr)\log(\rho^{o}(0,1))+\Bigl(\psi^{(1,0)}(2,2)+\psi^{(2,0)}(0,1;2,2)+\psi^{(2,0)}(1,1;2,2)\\
&+\psi^{(1,1)}(2,2;1,0)+\psi^{(1,1)}(2,2;1,1)\Bigr)\log(\rho^{o}(2,2))+\Bigl(\psi^{(0,1)}(1,0)+\psi^{(0,2)}(1,0;2,1)\\
&+\psi^{(0,2)}(1,0;2,2)+\psi^{(1,1)}(1,1;2,2)+\psi^{(1,1)}(2,2;1,0)\Bigr)\log(\rho^{x}(1,0))+\Bigl(\psi^{(0,1)}(2,2)\\
&+\psi^{(0,2)}(1,0;2,2)+\psi^{(0,2)}(1,1;2,2)+\psi^{(1,1)}(0,1;2,2)+\psi^{(1,1)}(1,1;2,2)\Bigr)\log(\rho^{x}(2,2))\\
&=\sum_{s\in V} f^{\text{ex}}(s)+f^{\text{id}}(s),
\end{align*}
with $f^{\text{ex}}(s)$ given by the new form in Eq.~\ref{Eq_density_new}. 
The calculation is analogous for the rotated cavity. Since the quadratic cavity can be embedded in this cavity, Lafuente and Cuesta's free energy density does not yield the correct functional in contrast to the other cavities determined in this chapter.
\end{appendices}

\end{document}